\documentclass[a4paper,10pt,aps,pre,twocolumn,amsmath,amssymb,longbibliography]{revtex4-1}

\usepackage{graphicx,color}
\usepackage[utf8]{inputenc}
\usepackage[colorlinks=true,citecolor=blue]{hyperref}
\usepackage{listings}
\usepackage{courier}
\usepackage{wasysym}

\newcommand{\ahum}[1]{``#1''}
\newcommand{\eq}[1]{Eq.~(\ref{#1})}
\newcommand{\fig}[1]{Fig.~\ref{#1}}

\newcommand{\olcite}[1]{Ref.~\cite{#1}}

\newcommand{\vk}{v_K}
\newcommand{\vc}{v_{\rm COM}}

\begin{document}

\title{Friction on incommensurate substrates: Role of anharmonicity and defects}

\author{S.~Amiri, C.A.~Volkert, and R.L.C.~Vink}

\affiliation{Institute of Materials Physics, Georg-August-Universität Göttingen, 
37073 Göttingen, Germany}

\date{\today}

\begin{abstract} We present Molecular Dynamics simulations of one- and two-dimensional bead-spring models sliding on incommensurate substrates. We investigate how sliding friction is affected by interaction anharmonicity and structural defects. In their absence, we confirm earlier findings, namely, that at special resonance sliding velocities, friction is maximal. When sliding {\it off-resonance}, partially thermalized states are possible, whereby only a small number of vibrational modes becomes excited, but whose kinetic energies are already Maxwell-Boltzmann distributed. Anharmonicity and defects typically destroy partial thermalization, and instead lead to full thermalization, implying much higher friction. For sliders with periodic boundaries, thermalization begins with vibrational modes whose spatial modulation is compatible with the incommensurate lattice. For a disc-shaped slider, modes corresponding to modulations compatible with the slider radius are initially the most dominant. By tuning the mechanical properties of the slider's edge, this effect can be controlled, resulting in significant changes in the sliding distance covered. \end{abstract}

\maketitle

\section{Introduction}

A sliding object generally loses its kinetic energy of forward motion and slows 
down due to friction. Understanding friction is still elusive in the majority of 
applications, for, if friction were understood, we would likely not be spending 
20\% of our energy consumption at trying to overcome 
it~\cite{10.1007/s40544-017-0183-5}. The loss of energy due to friction, by 
which one really means the conversion of useful energy (e.g.~forward kinetic 
motion) into less useful forms (e.g.~heat), can occur via many channels 
(phononic, electronic, magnetic, electrochemical, to name but a 
few~\cite{10.1007/978-3-662-04283-0, 10.1103/physrevb.59.11777, 
10.1103/physrevb.77.184105, 10.1103/physrevlett.101.137205, 
10.1103/physrevlett.112.055502}). One of these channels, the one we focus on in 
this paper using molecular dynamics simulations, is the phononic channel, i.e.~the loss of useful energy via the generation of internal lattice vibrations. The 
origins of phononic friction have been, and still are, actively 
discussed~\cite{10.1126/science.265.5176.1209, 10.1007/s11249-019-1247-7, 
10.1103/physrevb.54.8252, 10.1007/s11249-020-1268-2, 10.1007/s11249-020-1280-6, 
10.1103/physrevb.59.11777}. One view is that phonon scattering processes play a 
crucial role. Hence, materials with large anharmonicities in their interactions, 
or containing scattering centers such as structural defects, are likely to be 
more dissipative than materials without these 
properties~\cite{10.1126/science.265.5176.1209}.

Indeed, Molecular Dynamics (MD) simulations confirm that phonon scattering 
processes can be a major cause of energy dissipation under sliding, and the 
dissipation rate can be quantitatively related to phonon 
lifetimes~\cite{10.1063/1.5130705}. The results of \olcite{10.1063/1.5130705} 
were obtained using a 3D setup, consisting of an FCC copper substrate coated 
with a single layer of graphene, with a second graphene layer being dragged 
across the coated layer. The frictional properties could then be related to the 
lifetimes of phonons generated in the dragged graphene layer.

The setup of \olcite{10.1063/1.5130705} resembles a 2D Frenkel-Kontorova (FK) 
model~\cite{10.1007/978-3-662-10331-9}, such as studied in 
\olcite{10.1103/physreve.94.023001}. The essential difference is that, in the FK 
model, only in-plane particle motion of the sliding layer is retained, i.e.~the 
dynamics is strictly 2D. Nevertheless, even with this simplification,  
friction is higher when the particle interactions include some 
degree of anharmonicity, implying shorter phonon lifetimes~\footnote{See Fig.~8 
of Ref.~\cite{10.1103/physreve.94.023001}. The \ahum{vector} model variant, in 
which all higher-order terms in the particle interactions are retained, thus 
making it anharmonic, displays significantly higher friction.}. In another 
simulation study carried out by one of us~\cite{10.1103/physrevb.100.094305}, friction depended quite strongly on whether the substrate interactions 
were harmonic or not, being clearly correlated with the phonon lifetime (here: 
of phonons in the substrate, not in the slider, which in 
\olcite{10.1103/physrevb.100.094305} was a point particle). Further recent experimental studies also identify the phonon lifetime as key factor determining friction~\cite{schmidt2020switching, weber2020polaronic}.

The aim of this study is to further focus on the role of phonon scattering on sliding friction, considering interaction anharmonicity and structural defects. We will do so using the FK model, for which a large body of results is already available~\cite{10.1103/physrevlett.85.302, 10.1103/physreve.64.016601, 10.1103/physreve.94.023001}. We bring the system into an initial sliding state, then monitor how the kinetic energy of forward motion is converted into internal lattice vibrations, i.e.~heat. Our results apply to the FK model in its \ahum{floating} state, i.e.~in the absence of static friction. Such states can be realized using systems 
sliding on incommensurate substrates, and where the coupling to the substrate is weak (i.e.~below the Aubry transition~\cite{10.1088/0022-3719/16/9/005, 10.1016/0039-6028(93)91022-h, 10.1038/nmat4601}). The accepted view is that such systems, provided they are large enough, once brought into a sliding state, eventually thermalize, i.e.~with the initial kinetic energy of forward motion having been converted entirely into heat~\cite{10.1088/0953-8984/24/44/445009, 10.1103/physrevlett.85.302, 
10.1103/physreve.64.016601, 10.1103/physreve.94.023001}. However, depending on 
the precise system parameters (in particular, the initial sliding velocity~\cite{10.1103/physrevlett.85.302}), thermalization can be very slow. Possible experimental realizations could be small crystalline clusters (graphene flakes) sliding on incommensurate crystalline surfaces~\cite{10.1103/physrevlett.92.126101, 10.1209/0295-5075/95/66002}, clusters of Xe atoms sliding on Ag(111) 
substrates~\cite{10.1103/physrevlett.79.4798, 10.1088/0953-8984/24/44/445009}, 
or trapped ions in optical lattices~\cite{10.1038/nmat4601}.

\section{Model and Methods}

We consider FK models in $d=1$ (1D) and $d=2$ (2D) dimensions. For the 1D case, 
a large body of theoretical results is available (in particular 
\olcite{10.1103/physrevlett.85.302}) which help to guide the simulations, also 
in 2D. As stated previously, the extension of this work is to include additional 
phonon scattering processes, by means of anharmonicity and defects.

\subsection{1D FK model}

The 1D model considers a chain of $i=1,\ldots,N$ atoms, confined to move along a 
line, where periodic boundary conditions are applied. Each atom (uniform single atom 
mass $m$) in the chain is connected by springs to its two nearest neighbors. 
The energy of a single spring is given by
\begin{equation}
\label{eq:spr}
  u_{\rm spr} (r) = \sum_{n=2}^4 \epsilon \alpha_n (r/a-1)^n \quad,
\end{equation}
where $r$ denotes the distance between the two atoms participating in the bond, 
$a$ the equilibrium bond length, and where $\epsilon$ sets the energy scale. We 
will, in what follows, speak of harmonic and anharmonic systems. For the 
harmonic system, we use $\alpha_2=36,\alpha_3=\alpha_4=0$; for the anharmonic 
system $\alpha_2=36, \alpha_3=-252, \alpha_4=1113$. These parameters stem from a 
Taylor expansion of a $(12,6)$ Lenard-Jones potential around its minimum, with 
the minimum located at $r=a$, and well-depth $\epsilon$.
 
The total length of the system $L=aN$ such that, in the absence of any external 
fields, the chain groundstate energy equals zero. In addition to the mobile 
chain, an array of $M={\rm int}(gN)$ evenly-spaced static particles is 
distributed along the line, with $g=(1+\sqrt{5})/2$ the golden ratio, and where 
\ahum{int} means rounding down to the nearest integer. This choice ensures 
maximum incommensurability between the mobile chain and the static 
obstacles~\cite{10.1103/physreve.94.023001}, while remaining compatible with the 
periodic boundaries. The static obstacles interact with the mobile chain atoms 
via a soft pair potential of the form:
\begin{equation}
\label{eq:soft}
u_{\rm soft} (r) = \begin{cases}
\alpha \epsilon \left[1 + \cos \left(\frac{\pi r}{r_c}\right) \right] & r<r_c \\
0 & \mbox{otherwise,}
\end{cases}
\end{equation}
with $\alpha=0.3$ and $r_c = L/(2M)$. The use of an incommensurate static 
potential, and the relatively weak coupling between static layer and chain, 
ensures a \ahum{floating} state, whose friction is expected to be minimal (that is, we always stay below the Aubry transition~\cite{10.1088/0022-3719/16/9/005, 10.1016/0039-6028(93)91022-h, 10.1038/nmat4601}, i.e.~there is no static friction).

The undeformed chain (i.e.~with all the springs at their equilibrium length~$a$) 
is placed on the line containing the obstacles (a random uniform displacement is 
applied to all chain atoms, in order to sample different initial starting 
positions). At time $t=0$, the chain is \ahum{kicked} by assigning each chain 
atom the same velocity $\vk$ along the chain direction; the subsequent chain 
dynamics is then obtained by time-integrating the equations of motion in the 
micro-canonical ($NVE$) ensemble. Directly after kicking, the velocity of the 
chain center of mass equals $\vk$. However, due to the generation of vibrations in 
the chain (caused by collisions with the static obstacles, as well as, for the 
anharmonic chain, via internal phonon scattering) the velocity of the chain center 
of mass will typically decrease with time, i.e.~there is friction. We emphasize 
that no thermostat is applied in these simulations. Hence, results are 
completely free from thermostat-induced artifacts, which in friction simulations 
can be quite strong~\cite{10.1103/physrevb.100.094305, 
10.1103/physrevb.82.081401, 10.1007/s11249-012-9936-5}. The present approach 
thus facilitates an unbiased view into the origins of sliding 
friction, using what is arguably the optimally simplified \ahum{minimal} model.

\subsection{2D FK model}

\begin{figure}
\centering
\includegraphics[width=0.8\columnwidth]{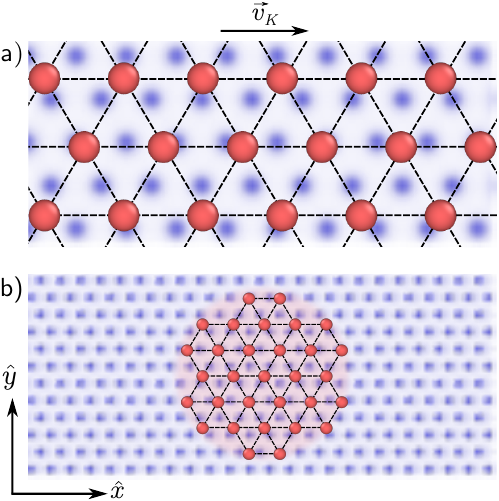}
\caption{\label{fig:2d_sketch} Schematics of the 2D FK model. Red circles 
represents the mobile atoms, which form a hexagonal lattice, whereby each atom 
is connected to its nearest neighbors by springs (dashed lines). The blue 
circles (blurred) represent the static obstacles which generate the potential 
energy landscape through which the mobile layer slides. The layer of mobile 
atoms is kicked with velocity $\vk$ along the $x$-axis, as indicated. We 
consider two geometries: a) sliding layer which is fully periodic in both 
dimensions, and b) a finite patch of sliding atoms (flake), approximately disc 
shaped.}
\end{figure}

The 2D model uses hexagonal lattices for both the mobile atoms and the static 
obstacles, with periodic boundaries applied in both directions. We consider two 
geometries, namely a fully periodic slider, and a finite patch (flake) of 
sliding atoms [\fig{fig:2d_sketch}]. For the fully periodic system, the mobile 
lattice contains $i=1,\ldots,N$ atoms, single atom mass $m$, each atom connected to its 
six nearest neighbors by springs. The aspect ratio of the lattice $L_y/L_x=\sqrt{3}/2$, with $L_i$ the length of the system in the direction $i \in x,y$. The single spring energy is given by \eq{eq:spr}, the spring rest length equals the lattice constant~$a$. The same definitions of harmonic and anharmonic bonds as used for the 1D chain are applied here as well. For the static incommensurate potential energy landscape, $M={\rm int}(g^2N)$ static particles are arranged on a second hexagonal lattice, using the same aspect ratio $L_y/L_x$ as the mobile lattice, where $g$ is the golden ratio. In this way, the ratio of 
lattice constants $a/a_c$ is as close as possible to $g$, where 
$a_c$ is the lattice constant of the static layer, ensuring maximum 
incommensurability. The interaction between the static obstacles and the mobile 
atoms is again of the form of \eq{eq:soft}, with $\alpha=0.3$ and $r_c=a_c/2$. 
The 2D hexagonal layer of mobile atoms is placed inside the static potential 
energy landscape generated by the obstacles, oriented as shown in 
\fig{fig:2d_sketch}(a). At time $t=0$, all the atoms in the mobile layer are 
\ahum{kicked}, by assigning them the velocity $\vk$ in the $\hat{x}$ direction 
(prior to kicking, the mobile lattice is given a random uniform 2D displacement, 
in order to sample different starting positions). We emphasize that the model is purely 2D, i.e.~the mobile atoms can move in the $\hat{x}$ and $\hat{y}$ directions only. Immediately after kicking, the motion is entirely in the $\hat{x}$ direction, but soon thereafter, due to collisions with the static obstacles, also motion in the $\hat{y}$ direction develops. For the flake, a finite portion of the hexagonal lattice is retained, 
keeping only those atoms inside a specified radius of some central reference 
atom [\fig{fig:2d_sketch}(b)]. Most of the flake atoms will be six-fold 
coordinated, except for those on the edge, which have missing bonds. The flake 
is oriented with respect to the static lattice in the same way as the fully 
periodic slider. The static lattice is chosen large enough to fully encompass 
the flake, such that periodic boundary conditions may safely be applied.

\subsection{Units}

For both the 1D and 2D model, length is expressed in units of the equilibrium 
lattice constant $a$, energy in units of $\epsilon$, particle mass in units of 
some reference mass $m^\star$, and temperature in units of $\epsilon/k_B$, with 
$k_B$ the Boltzmann constant. This implies time unit $[t] = \sqrt{m^\star a^2 / 
\epsilon} \sim 2.6 \, \rm ps$ assuming a sliding layer consisting of C-atoms ($a 
\approx 3.8 \, \rm \AA$, $m^\star \approx 12 \, \rm u$, $\epsilon \approx 2.76 
\, \rm meV/atom$~\cite{10.1039/c0cp02614j}).

\subsection{Eigenmodes}

To analyse the vibrational excitations in the mobile lattice induced during 
sliding, we use the language of eigenmodes, which has proven to be useful in 
other studies also~\cite{10.3762/bjnano.8.218}. For a system of $i=1,\ldots,N$ 
particles, there are $k=1,\ldots,dN$ eigenmodes, with $d=1,2$ the spatial 
dimension of the problem at hand. The eigenmodes follow in the usual way from 
the (mass-weighted) hessian, $H_{\mu\nu} = \frac{1}{\sqrt{m_\mu m_\nu}} 
\frac{\partial^2 E}{\partial \mu \partial \nu}$, with $E$ the total spring 
energy of the system given by \eq{eq:spr}, and with the derivatives evaluated 
with the sliding atoms in their perfect equilibrium lattice positions (of 
course, when computing the hessian, the interaction with the static particles is 
excluded). The labels $\mu,\nu$ refer to the set of all Cartesian coordinates of 
the particles, $m_{\mu,\nu}$ being the associated particle mass. The hessian is a $dN 
\times dN$ matrix, but most elements are zero, since the particles interact only 
with nearest neighbors. Upon diagonalizaton of the hessian, a set of 
eigenvectors $\vec{\xi}_k$ is obtained, each one with an associated 
eigenfrequency $\omega_k^2$. For the 1D chain, there is exactly one mode with 
zero eigenfrequency, corresponding to a global translation of the chain along 
the $x$-axis. For the 2D sliding layer, there will always be at least two zero 
frequency modes, corresponding to global translations in the two lateral 
directions. In addition, if the 2D layer is a finite patch, there will also be a 
third zero frequency mode, corresponding to a global rotation. For lattices with perfect translational symmetry (i.e.~fully periodic, defect-free crystals), one can assign a wavevector to each eigenmode, then corresponding to a true phonon.

During the sliding simulations, we record, for each particle, the displacement 
$\vec{u}_i (t)$ from the initial (perfect lattice) position, and velocity 
$\vec{v}_i (t)$, both as functions of time~$t$ (for the 1D chain, these 
quantities are scalars; for the 2D sliding layer, they are 2D vectors). From 
these, we define the kinetic energy of the $k$-th eigenmode as:
\begin{equation}
\label{eq:kin}
K_k (t) = \frac{1}{2} \left( \sum_{i=1}^N \sqrt{m_i} \, 
\vec{v}_i (t) \cdot \vec{\xi}_{k,i} \right)^2 \quad,
\end{equation}
with the sum over all particles, $m_i$ the mass of particle~$i$, and 
$\vec{\xi}_{k,i}$ the sub-vector of the full eigenvector $\vec{\xi}_k$, 
containing only the components of particle~$i$. Defined in this way, one 
consistently has
\begin{equation}
\label{eq:chk}
 E_{\rm kin} = \sum_{i=1}^N \frac{m_i \vec{v}_i^2}{2}
 = \sum_{k=1}^{dN} K_k \quad,
\end{equation}
which holds exactly (for both harmonic and anharmonic systems).

\section{Results}

All our MD results were obtained with LAMMPS~\cite{10.1006/jcph.1995.1039}; 
implementation details are provided in the Appendix.

\subsection{1D chain}

\begin{figure}
\begin{center}
\includegraphics[width=\columnwidth]{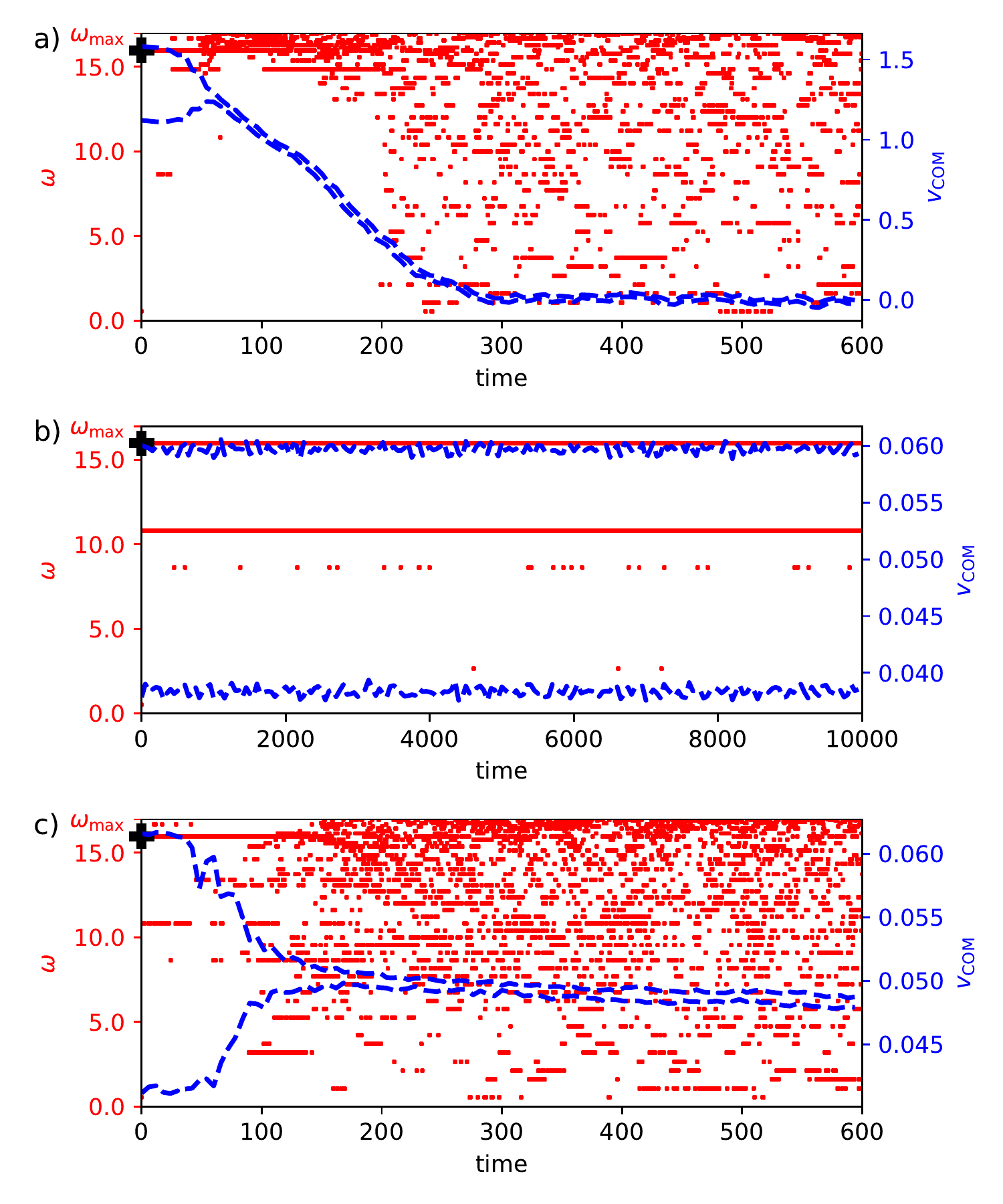}
\caption{\label{fig1} 1D sliding chain results, showing time evolution directly 
after kicking with velocity $\vk$ (results averaged over 20 trajectories, each 
with different initial position of the slider). The dashed blue curves show the 
lower and upper {\it envelope} of the chain center of mass velocity $\vc$. The 
red dots indicate, for each time step, the frequency of the kinetically most 
active mode. Symbol~$+$ on the vertical axes indicates $\omega^\star$; the 
maximum mode frequency $\omega_{\rm max} \approx 16.97$. (a) Harmonic chain 
kicked with the resonance velocity $\vk^\star$. The decay of $\vc$ sets in 
rapidly after kicking. (b) Harmonic chain kicked with $\vk=0.05 \ll \vk^\star$. 
In this case, $\vc$ oscillates between a low and high value, but there is no 
decay. Only a small subset of available modes reveals noticeable activity. (c) 
Same as b), but using {\it anharmonic} interactions. The decay of $\vc$, and 
subsequent thermalization, commence rapidly after kicking.}
\end{center}
\end{figure}

We consider a chain with $N=100$ particles, periodic boundaries, unit particle mass 
$m=1$. Unless stated otherwise, the bond interaction of 
\eq{eq:spr} is harmonic. For these parameters, the frictional behavior is well 
understood~\cite{10.1103/physrevlett.85.302}. The static obstacles induce a 
spatial modulation of wavenumber $k^\star = 2\pi/a_c$ in the 
chain~\cite{10.1103/physrevb.60.6522}, with $a_c$ the lattice spacing of the 
static obstacles. The chain center of mass motion thus couples to the chain 
internal vibrations via the mode $k^\star$; the associated vibrational frequency 
follows from the dispersion relation $\omega^\star = 2 
\sqrt{2\alpha_2\epsilon/m} |\sin(ak^\star/2)|$. When kicked with velocity $\vk$, 
chain atoms \ahum{hit} the obstacles with the washboard frequency $\Omega = 
\vk/a_c$. Friction arises when a resonance is created, $2\pi\Omega \sim 
\omega^\star$. Via a cascade of couplings between $k^\star$ and the other 
vibrational modes in the chain, the kinetic energy of the center of mass forward 
motion is transferred, via $k^\star$, to the entire population of chain 
vibrational modes, thereby converted into heat.

For our model parameters $\omega^\star \approx 15.97$, the corresponding 
resonance kick velocity $\vk^\star \approx 1.58$. When the chain is kicked with 
$\vk^\star$, the chain center of mass velocity decays rapidly with time, 
i.e.~friction is high [\fig{fig1}(a)]. In contrast, using $\vk=0.05$, which is 
far below resonance, $\vc$ oscillates between a low and high value, but there is 
no sign of any decay, i.e.~friction is low [\fig{fig1}(b)]. Also indicated in 
\fig{fig1} is the frequency of the kinetically most active mode as a function of 
time, defined as the mode having the highest value of $K_k$, as given by 
\eq{eq:kin}. In the low-friction state, \fig{fig1}(b), only a few modes are 
active. These are the modes $k^\star$, as well as some of the {\it higher 
harmonics}, corresponding to wavenumber $nk^\star$, with $n$ a positive integer. 
In the high-friction state, \fig{fig1}(a), at very early times, we also observe 
that activity is concentrated around $k^\star$, but soon spreads to all modes, 
reminiscent of a system in thermal equilibrium (the signal $K_k$ then 
essentially being a random variable).

The low-friction state of \fig{fig1}(b) can persist because, being {\it 
off-resonance}, the coupling of $k^\star$ to other vibrational modes is weak, 
{\it and} because the chain interaction is harmonic (i.e.~no scattering between 
modes). In such a highly de-coupled system, the transfer of energy between modes 
is severely hampered, meaning that thermalization (i.e.~generation of heat) 
cannot occur, which explains why friction is low. Indeed, by using {\it 
anharmonic} bonds, which enable mode scattering thereby assisting 
thermalization, the second condition no longer holds, and the low-friction state 
is no longer observed [\fig{fig1}(c)].

\begin{figure}
\begin{center}
\includegraphics[width=\columnwidth]{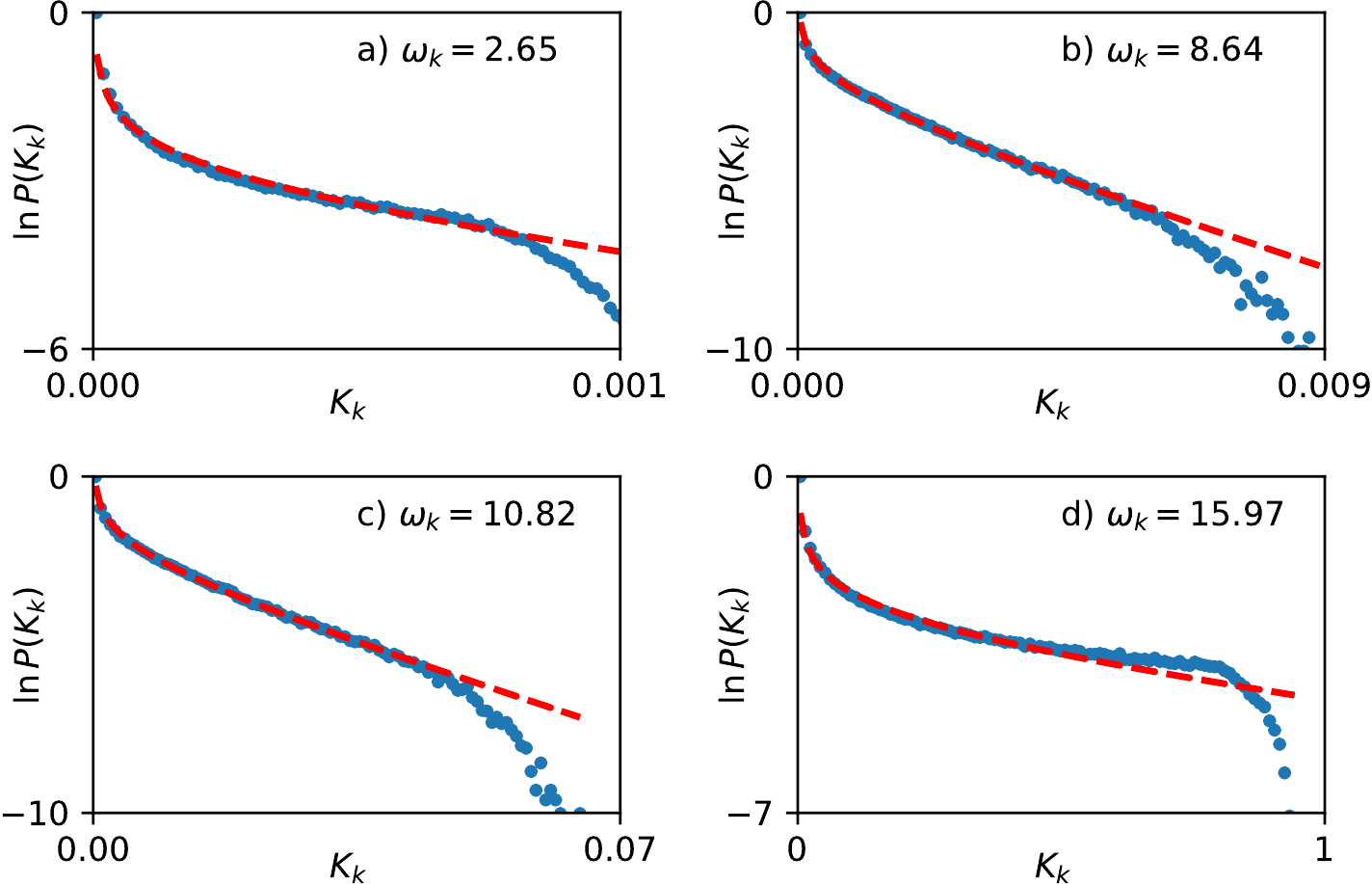}
\caption{\label{fig2} 1D {\it harmonic} chain results, showing the logarithm of 
the distribution of mode kinetic energies, obtained during sliding in the 
low-friction state of \fig{fig1}(b). Results are shown for the four most active 
modes, with frequencies $\omega_k$ as indicated. The dashed lines show fits to 
\eq{eq:pk}, which are two-parameter fits, one of them being the mode 
temperature~$T_k$.}
\end{center}
\end{figure}

\begin{figure}
\begin{center}
\includegraphics[width=\columnwidth]{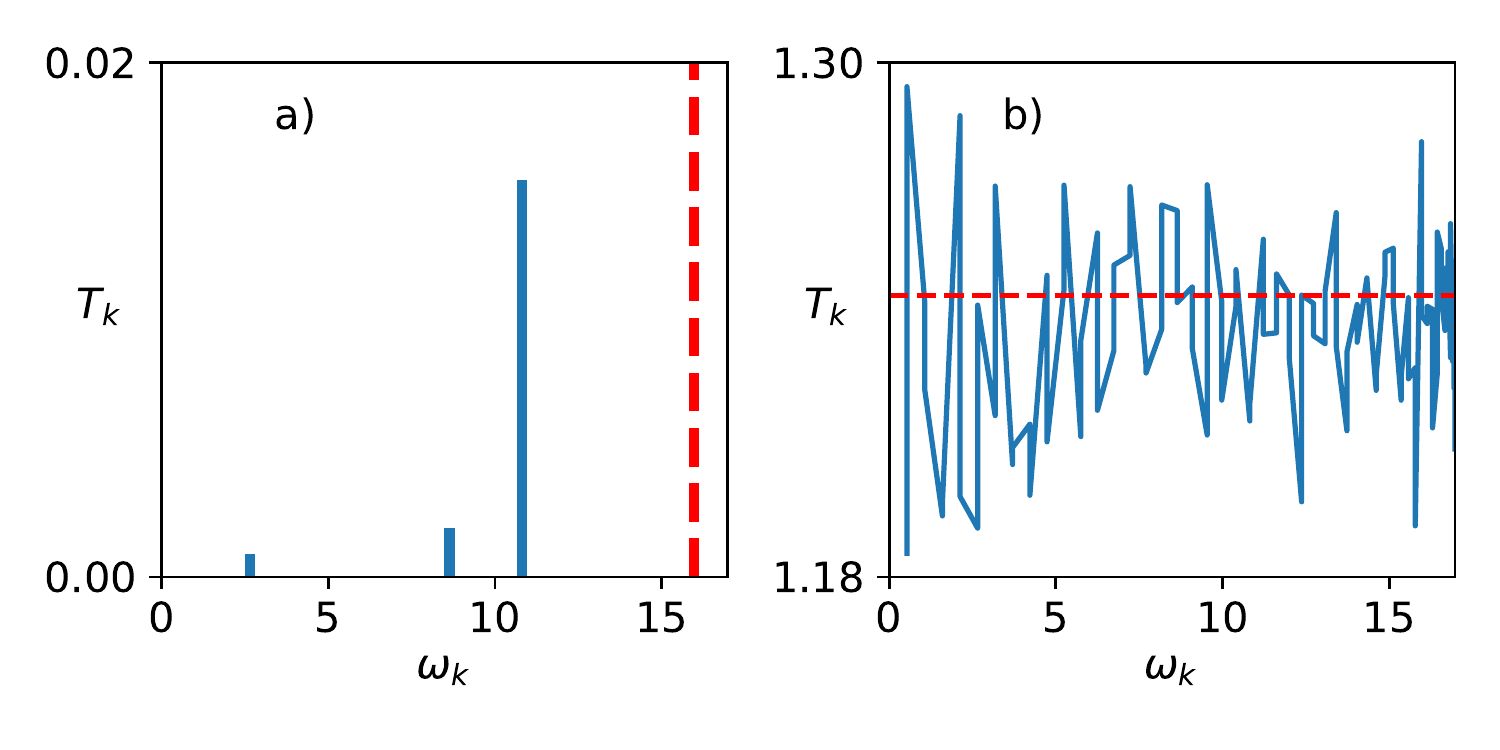}
\caption{\label{fig3} 1D {\it harmonic} chain results, showing mode temperatures 
$T_k$ obtained by fitting to \eq{eq:pk}, as function of the mode frequency 
$\omega_k$. a) As obtained in the low-friction state of \fig{fig1}(b). Note: 
$T_k$ for the principal mode $k^\star$ (dashed line) far exceeds the vertical 
range of the graph. b) As obtained in the long-time limit after kicking with the 
resonance velocity $\vk^\star$; dashed line marks the equipartition 
temperature.}
\end{center}
\end{figure}

Next, we address thermalization. The low-friction state of \fig{fig1}(b) is not 
thermalized, since only a small subset of modes is active. Nevertheless, 
precursors to thermalization are already present. To see this, we consider the 
four most active modes of \fig{fig1}(b), whose wavenumbers are $k=nk^\star$, 
with $n=1,2,3,5$ (i.e.~the fundamental mode, and some of the leading higher 
harmonics, excluding $n=4$, which showed very little activity). The respective 
vibrational frequencies are $\omega_k=15.97; 10.82; 8.64; 2.65$. For these 
modes, while sliding in the low-friction state, the distribution (histogram) of 
observed kinetic energy values $K_k$ is recorded. In a perfectly thermalized 
system, these values are Maxwell-Boltzmann distributed, $P_{\rm th} \propto 
e^{-K_k / k_BT_k}$, with $T_k$ the mode temperature, and $k_B$ the Boltzmann 
constant. In a perfectly coherent state, $K_k$ as function of time is strictly 
harmonic, at {\it twice} the mode frequency, in which case the distribution 
takes the form $P_{\rm coh} \propto K_k^{-1/2}$, valid in the limit of small 
$K_k$ (Appendix). However, the low-friction state considered here is neither 
fully thermalized nor coherent, and so we expect a {\it hybrid} form:
\begin{equation} 
\label{eq:pk}
 P(K_k) = P_{\rm th} \cdot P_{\rm coh} \propto 
 e^{-K_k / k_BT_k}/\sqrt{K_k} \quad.
\end{equation}
We test the validity of \eq{eq:pk} in \fig{fig2}, for each of the four most 
active modes. The dashed curves are fits using \eq{eq:pk}. Overall, the fits 
capture the data well. In all cases, agreement breaks down at large values of 
$K_k$, since, on the one hand, \eq{eq:pk} is a small $K_k$ approximation, but, 
more importantly, due to bad statistics (large values of $K_k$ are exponentially 
suppressed by the Maxwell-Boltzmann factor, so these values do not appear very 
often in the simulation time series). 

We repeat the analysis of \fig{fig2} for all modes $k$ in the chain, to obtain 
the mode temperatures $T_k$. In the low-friction state, there are just a few 
active modes with finite temperature, inside a background of frozen modes 
[\fig{fig3}(a)]. The partial thermalized character of the low-friction state is 
clearly visible: While individual modes already have energy distributions 
conforming to Maxwell-Boltzmann, the corresponding temperatures between modes 
are very different. \fig{fig3}(b) shows the mode temperatures $T_k$ obtained 
after kicking with the resonance velocity $\vk^\star$, in the long-time limit 
where $\vc \sim 0$. We now observe a much more homogeneous temperature 
distribution, all modes having essentially the same temperature, showing that 
the chain has fully thermalized. For the {\it harmonic} chain in thermal 
equilibrium, equipartition should hold, i.e.~the initial kinetic energy of the 
kick ($K_{\rm in}=mN\vk^2/2$) should be equally divided over all system degrees 
of freedom (${\rm ndof}=2dN$, with $d=1$ the spatial dimension; factor two 
counts position and momentum degrees of freedom). For the harmonic chain in 
equilibrium, $k_BT/2 = K_{\rm in} / {\rm ndof}$, implying $T \approx 1.246$ in 
our units, which \fig{fig3}(b) confirms.

\subsection{2D hexagonal layer}

\begin{figure}
\begin{center}
\includegraphics[width=\columnwidth]{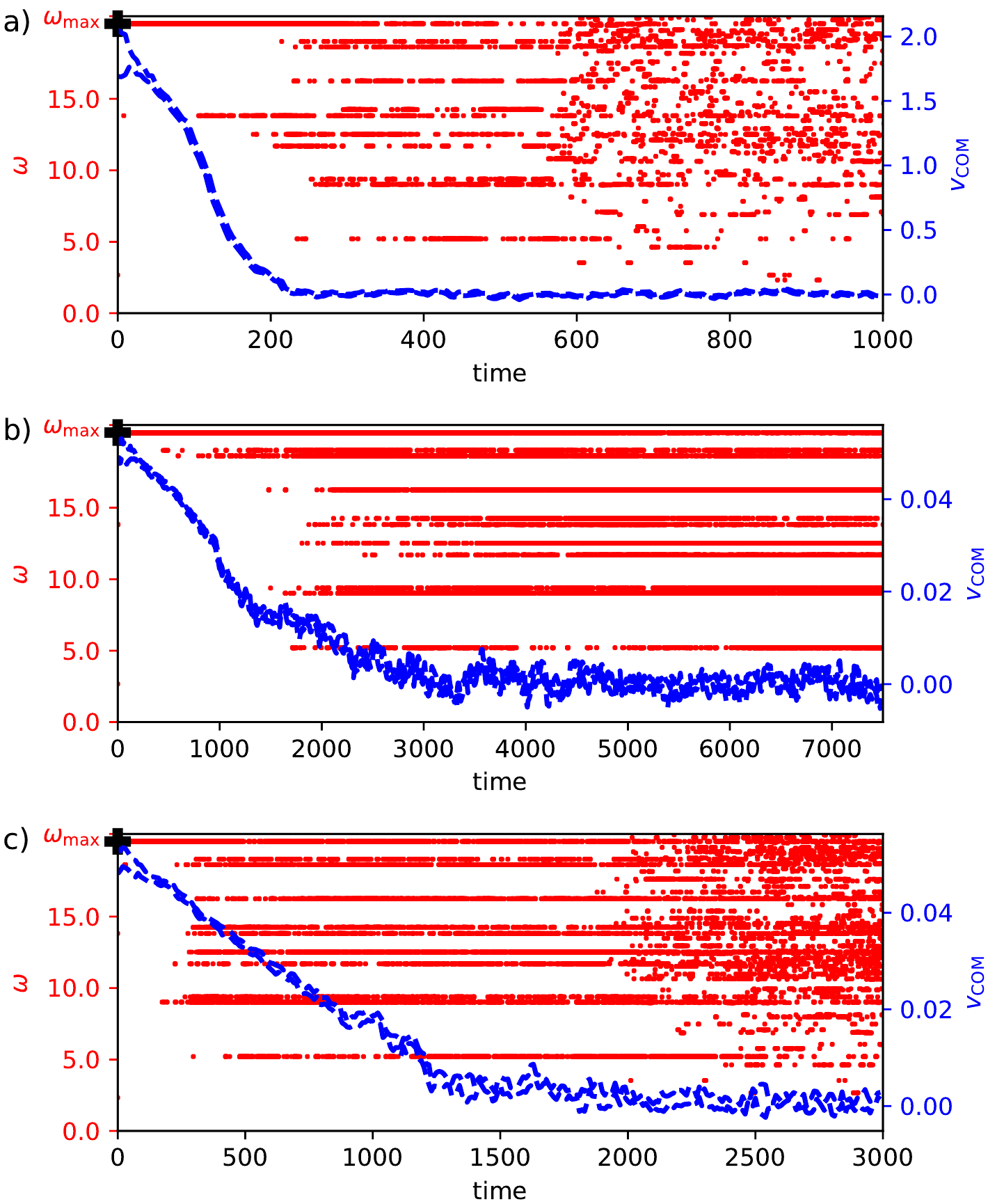}
\caption{\label{fig:2d_har_anhar} Sliding behavior of the 2D fully periodic 
slider (results are averaged over 20 different initial positions of the slider). The representation of the data is the same as in \fig{fig1}. Symbol $+$ indicates the frequency $\omega^\star$ of the dominant mode $k^\star$, the maximum possible mode 
frequency $\omega_{\rm max} \approx 20.78$. Results are shown for a) harmonic 
interactions at the resonance kick velocity $\vk^\star$, b) harmonic 
interactions at $\vk=0.05$, and c) {\it anharmonic} interactions at $\vk = 
0.05$.}
\end{center}
\end{figure}

\subsubsection{2D fully periodic slider without defects}

We first consider a 2D sliding layer with full periodic boundary conditions, 
i.e.~in the absence of any free edges or other defects [\fig{fig:2d_sketch}(a)]. 
A layer of $N=196$ mobile atoms, unit particle mass $m=1$, is \ahum{kicked} 
at time $t=0$ with velocity $\vk$ in the $\hat{x}$-direction. For this value of 
$N$, the lattice constant of the static obstacles $a_c=7a/11$. In analogy with 
the 1D chain, we assume that the static obstacles induce a spatial modulation of 
wavevector {\it magnitude} $k^\star = 2\pi/s$, with $s=a_c \sin 60^{\rm o}$ the 
spacing between closed-packed rows of obstacles, see \fig{fig:2d_sketch}(a). As for the {\it direction} and {\it polarization}, we assume that longitudinal modes propagating at $\pm 30^{\rm o}$ relative to $\hat{x}$ will be the dominant excitation. The corresponding vibrational frequency $\omega^\star \approx 20.26$, which follows from the dispersion relation (Appendix). For sliding in the $\hat{x}$-direction, the washboard 
frequency $\Omega=\vk/a_c$, implying resonance kick velocity $\vk^\star \approx 2.05$.

For the slider with harmonic bonds, the decay of $\vc$ with time at the 
resonance kick velocity $\vk^\star$ is shown in \fig{fig:2d_har_anhar}(a), while 
\fig{fig:2d_har_anhar}(b) shows the result for $\vk=0.05$, i.e.~far below 
resonance. In agreement with the 1D chain, the decay is most rapid at resonance, 
i.e.~friction is highest there. In addition, strong initial activity of the mode 
$k^\star$ is observed, confirming the above assumption that longitudinal modes 
propagating at $\pm 30^{\rm o}$ couple most strongly to the center of mass 
motion (the other plateaus visible in \fig{fig:2d_har_anhar} correspond to 
higher harmonics $nk^\star$). Regarding as to how the energy gets distributed 
over the vibrational modes, there is an important qualitative difference with 
the 1D chain. In 2D, see \fig{fig:2d_har_anhar}(b), a state is observed where 
$\vc \sim 0$, while the vibrational modes are still far from thermal 
equilibrium. This state is analogous to the low-friction state of \fig{fig1}(b), 
the crucial difference being that, in 2D, $\vc \sim 0$, i.e.~the system is no 
longer sliding. Repeating the simulation using $\vk=0.05$ and {\it anharmonic} 
bonds, \fig{fig1}(c), we observe a slightly more rapid decay of $\vc$ compared 
to the harmonic case at the same kick velocity, but this time the system fully 
thermalizes, i.e.~all modes become active.

\begin{figure}
\centering
\includegraphics[width=\columnwidth]{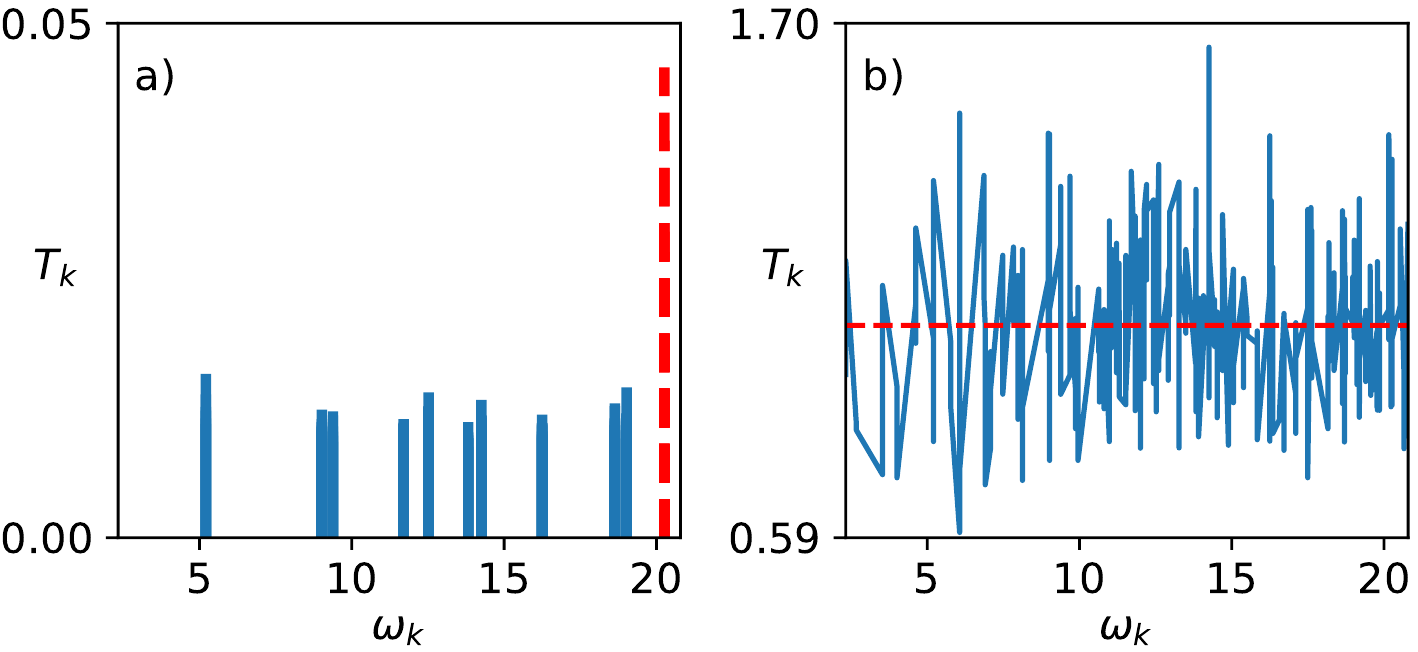}
\caption{\label{fig:2d_Tk} The analogue of \fig{fig3}, but for the 2D periodic 
slider with harmonic bonds. a) As obtained in the partially thermalized state of 
\fig{fig:2d_har_anhar}(b), where the slider was kicked with an {\it 
off-resonance} velocity $\vk=0.05$. b) As obtained in the long-time limit of 
\fig{fig:2d_har_anhar}(a), where the slider was kicked with the resonance 
velocity $\vk^\star$. In this case, there is full thermalization.}
\end{figure}

For the harmonic sliders, we still verify the degree of thermalization. For the 
slider in the partially thermalized state, \fig{fig:2d_har_anhar}(b), mode 
activity is mainly restricted to $k^\star$ and the higher harmonics. As in the 1D case, the kinetic energy distributions of these modes already appear thermalized, i.e.~well 
described by \eq{eq:pk}. In \fig{fig:2d_Tk}(a), we plot the corresponding mode 
temperatures, which reveals many frozen modes ($T_k \sim 0$), and a number of 
active modes ($T_K>0$), confirming that the state is indeed partially thermalized 
(for a fully thermalized state, $T_k$ should be the same for all modes). 
Compared to the analogous 1D case, \fig{fig3}(a), we find that in 2D the mode temperature is more homogeneous (with the exception of $\omega^\star$, the active modes have similar temperatures). In \fig{fig:2d_Tk}(b), we show the mode temperatures for the slider kicked with the resonance velocity $\vk^\star$, i.e.~corresponding to \fig{fig:2d_har_anhar}(a), in the long-time limit. In this case, the system fully thermalizes, all mode temperatures being the same. Note that equipartition is obeyed quite well, $k_B T_{\rm eq} = m(\vk^\star)^2/4 \approx 1.05$, as indicated by the dashed horizontal line. For the anharmonic slider, \fig{fig:2d_har_anhar}(c), the system also fully thermalizes, but this comes as no surprise, due to the enhanced phonon scattering induced by anharmonicity (result therefore not shown).

\subsubsection{2D slider with defects}

\begin{figure}
\begin{center}
\includegraphics[width=\columnwidth]{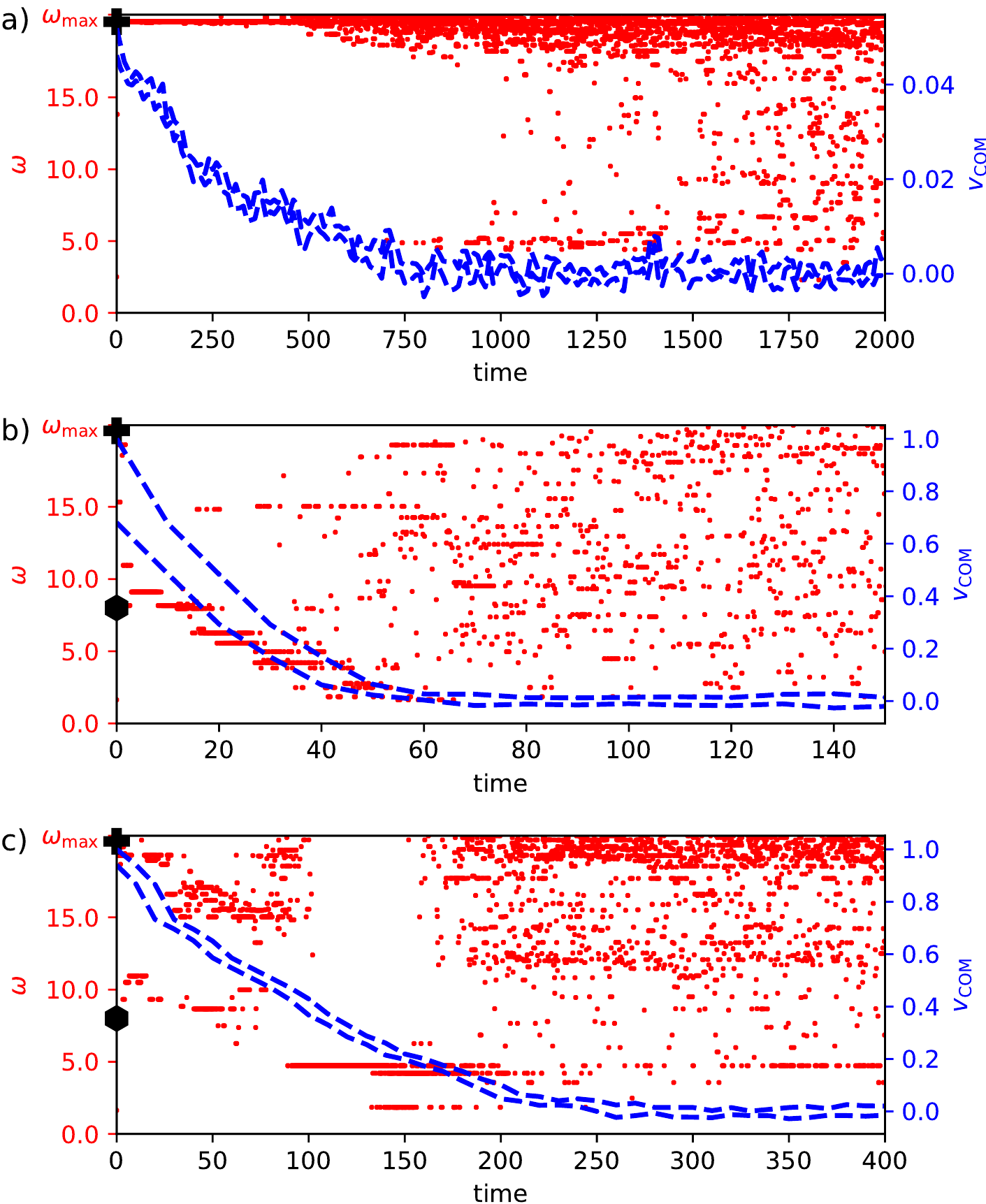}
\caption{\label{fig:2d_def_anhr_fl} 2D {\it harmonic} sliders containing defects (results are averaged over 20 different initial positions of the slider). The representation of the data is the same as in \fig{fig1}. Symbol $+$ indicates $\omega^\star$ of the modulation $k^\star$ induced by the static obstacles; symbol $\varhexagon$ indicates $\omega_R$ induced by the slider radius. a) Fully periodic slider with a fraction 2\% of randomly selected missing bonds, at kick velocity $\vk=0.05$. b) Sliding patch (flake) kicked with velocity $\vk=1$. c) Same as b), but for a slider with rigid edge.}
\end{center}
\end{figure}

We now investigate the role of lattice defects on the sliding behavior, 
considering bond and edge defects. For the bond defects, we remain with the 
fully periodic slider considered previously (same number of particles $N=196$; unit particle mass $m=1$) but with a fraction 2\% of randomly selected bonds 
removed from the lattice (we checked that, for this low fraction, the lattice 
remains a single connected entity, i.e.~there are no isolated atoms without any 
bonds). \fig{fig:2d_def_anhr_fl}(a) shows the corresponding sliding behavior, 
using harmonic interactions and kick velocity $\vk=0.05$, to be compared to the 
corresponding defect-free case of \fig{fig:2d_har_anhar}(b). The difference is 
striking: Whereas the defect-free slider did not thermalize, the presence of 
just a small number of defects strongly promotes thermalization, implying a much 
higher friction. Mode activity at short times is still concentrated around the dominant mode $k^\star$, but enhanced activity of the higher harmonics is no longer observed. Apparently, the presence of just a small number of defects is sufficient to destroy the coupling between $k^\star$ and its higher harmonics.

To study the influence of edge defects, we consider a disc-shaped slider 
(flake), see \fig{fig:2d_sketch}(b). The flake contains $N=199$ particles, 
i.e.~comparable to the fully periodic system; unit particle mass $m=1$. The edge of the slider provides an additional source of phonon scattering, which dramatically reduces sliding. In fact, at low kick velocity, 
$\vk=0.05$, the flake refuses to slide at all, merely a damped rocking motion of 
the center of mass is observed, irrespective of whether harmonic or anharmonic 
bonds are used. The damping is very strong, and the system thermalizes rapidly 
(results not shown). To observe any sliding at all, higher kick velocities are 
required. In \fig{fig:2d_def_anhr_fl}(b), we show results for $\vk=1$, using 
harmonic interactions. We find that the system thermalizes extremely rapidly, 
even faster than the fully periodic slider at the resonance velocity $\vk^\star$ [cf.~\fig{fig:2d_har_anhar}(a)]. Note also that initial mode activity is no longer concentrated at $\omega^\star \approx 20.26$ induced by the static obstacles, but instead at a much lower frequency. For the flake, the dominant spatial modulation is set by the flake radius, $k_R \sim 2\pi/R$, where $R \sim 6.9a$ presently. From the dispersion relation, and assuming longitudinal modes at $\pm 30^{\rm o}$ still dominate, this leads to a vibrational frequency $\omega_R \sim 8$, which is indeed rather close to the frequency where initially much activity is observed, see \fig{fig:2d_def_anhr_fl}(b). By making the edge of the slider infinitely stiff (i.e.~treat the edge as a rigid object, while time-integrating the internal particles as before, some of which with bonds to the, now rigid, edge) one can reduce the spatial modulation $k_R$. In this case, still kicking with velocity $\vk=1$, the decay of $\vc$ can be postponed, see \fig{fig:2d_def_anhr_fl}(c). Note that, by reducing the modulation $k_R$, the modulation $k^\star$ becomes visible again, leading to initial mode activity at both frequencies, $\omega_R$ and $\omega^\star$, simultaneously.

\begin{figure}
\begin{center}
\includegraphics[width=\columnwidth]{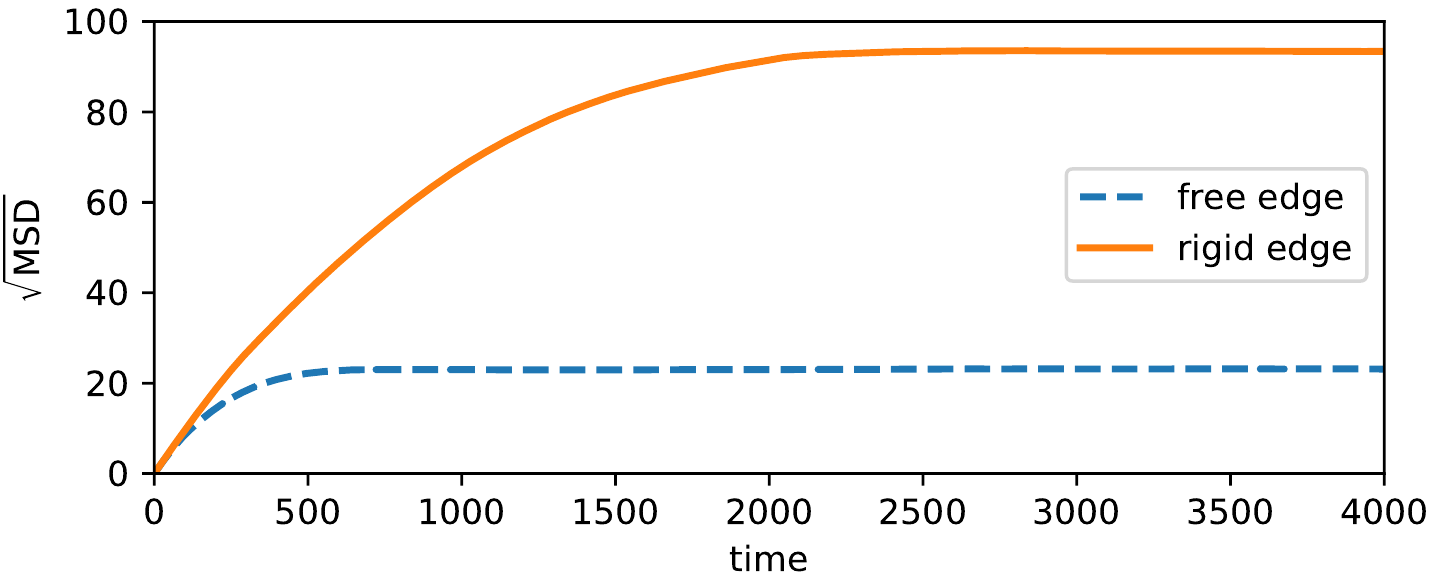}
\caption{\label{fig:msd} Total sliding distance versus time, as expressed via the mean-squared-displacement, for the flake with free and rigid edge. All interactions are harmonic, the kick velocity $\vk=1$.}
\end{center}
\end{figure}

As possible control tactic to reduce friction, the results of the sliding flake suggest optimizing the mechanical properties of the slider, in order to reduce the spatial modulation $k_R$ induced by the finite system size. As shown above, one way this may be achieved is to make the edge of the slider as stiff as possible (rigid). This results in a significant increase of the covered sliding distance, $s = \sqrt{\rm MSD}$, where MSD is the mean-squared-displacment of the slider atoms, as measured from the time of the 
kick [\fig{fig:msd}]. As the figure shows, the slider with the rigid edge slides 
roughly five times further.

\section{Conclusions}

We have investigated the sliding behavior of 1D and 2D bead-spring models on incommensurate substrates, in the \ahum{floating} state, i.e.~below the Aubry transition. For the 1D system, our results are fully consistent with the theoretical predictions of \olcite{10.1103/physrevlett.85.302}. For harmonic chain interactions, friction is highest when the washboard frequency corresponding to the kick velocity $\vk$ resonates with the dominant vibrational mode induced by the incommensurate substrate. For $\vk$ chosen {\it off-resonance}, a low-friction state is possible, where the system slides seemingly indefinitely, with only a small subset of the system vibrational modes showing any activity. As was already known~\cite{10.1103/physrevlett.85.302}, the low-friction state can only survive in sufficiently small systems, such that the vibrational spectrum remains discrete. One insight of this work is that, in addition, the interactions must be sufficiently harmonic, since anharmonicity will also destroy the low-friction state. A further insight is that the low-friction state is already partially thermalized, the kinetic energies of the active vibrational modes being well described by a modified Maxwell-Boltzmann factor. Thermal fluctuations (i.e.~randomness) are thus already present, which could imply that the low-friction state unavoidably has a finite lifetime.

In 2D, for the system size considered here, a low-friction state where the system slides indefinitely, was not observed. This is consistent with \olcite{10.1103/physreve.94.023001}, where it was also found that true 2D models typically equilibrate, rather than slide, even when the system size is small. Instead, we find that partially thermalized states are possible, with only a few active vibrational modes, but where the center of mass velocity has already decayed to zero. These partially thermalized states can occur when the system is kicked with an {\it off-resonance} velocity, and for harmonic interactions. In line with the 1D system, the kinetic energies of the active modes are Maxwell-Boltzmann distributed, so thermal fluctuations already manifest themselves. For anharmonic interactions, the partially thermalized state is also observed, but here its duration is very brief, full thermalization setting in quickly.

For both the 1D and 2D periodic sliders, but without defects, the vibrational modes that initially get excited correspond to the dominant spatial modulation induced by the incommensurate substrate and higher harmonics. In the presence of point defects, but still with periodic boundaries, only the dominant spatial modulation gets excited, the coupling to higher harmonics then appears lost. An even more striking effect is observed for sliders with edges: In this case, initial mode activity may instead commence at spatial modulations corresponding to the radius of the slider, the degree of which is controlled by the edge stiffness.

Regarding applications, for which a true low-friction state with indefinite 
sliding is likely of most interest, the sobering news is that the system 
parameters must be very carefully chosen: highly harmonic interactions, small 
systems, defect-free. However, even if these conditions cannot be perfectly met, 
there is still the option to reduce friction, for example by tuning the 
mechanical properties of the slider edge. Interestingly, a recent 
publication~\cite{10.1039/c4nr06521b} also identifies the importance of edges 
concerning {\it static} friction, so their relevance seems to extend beyond the 
purely {\it dynamic} scenarios considered here.

\acknowledgments

This work was funded by the Deutsche Forschungsgemeinschaft (DFG, German 
Research Foundation) -- 217133147/SFB 1073, project A01.

\bibliography{MISC,AUTODOI}

\begin{thebibliography}{35}%
\makeatletter
\providecommand \@ifxundefined [1]{%
 \@ifx{#1\undefined}
}%
\providecommand \@ifnum [1]{%
 \ifnum #1\expandafter \@firstoftwo
 \else \expandafter \@secondoftwo
 \fi
}%
\providecommand \@ifx [1]{%
 \ifx #1\expandafter \@firstoftwo
 \else \expandafter \@secondoftwo
 \fi
}%
\providecommand \natexlab [1]{#1}%
\providecommand \enquote  [1]{``#1''}%
\providecommand \bibnamefont  [1]{#1}%
\providecommand \bibfnamefont [1]{#1}%
\providecommand \citenamefont [1]{#1}%
\providecommand \href@noop [0]{\@secondoftwo}%
\providecommand \href [0]{\begingroup \@sanitize@url \@href}%
\providecommand \@href[1]{\@@startlink{#1}\@@href}%
\providecommand \@@href[1]{\endgroup#1\@@endlink}%
\providecommand \@sanitize@url [0]{\catcode `\\12\catcode `\$12\catcode
  `\&12\catcode `\#12\catcode `\^12\catcode `\_12\catcode `\%12\relax}%
\providecommand \@@startlink[1]{}%
\providecommand \@@endlink[0]{}%
\providecommand \url  [0]{\begingroup\@sanitize@url \@url }%
\providecommand \@url [1]{\endgroup\@href {#1}{\urlprefix }}%
\providecommand \urlprefix  [0]{URL }%
\providecommand \Eprint [0]{\href }%
\providecommand \doibase [0]{http://dx.doi.org/}%
\providecommand \selectlanguage [0]{\@gobble}%
\providecommand \bibinfo  [0]{\@secondoftwo}%
\providecommand \bibfield  [0]{\@secondoftwo}%
\providecommand \translation [1]{[#1]}%
\providecommand \BibitemOpen [0]{}%
\providecommand \bibitemStop [0]{}%
\providecommand \bibitemNoStop [0]{.\EOS\space}%
\providecommand \EOS [0]{\spacefactor3000\relax}%
\providecommand \BibitemShut  [1]{\csname bibitem#1\endcsname}%
\let\auto@bib@innerbib\@empty
\bibitem [{\citenamefont {Holmberg}\ and\ \citenamefont
  {Erdemir}(2017)}]{10.1007/s40544-017-0183-5}%
  \BibitemOpen
  \bibfield  {author} {\bibinfo {author} {\bibfnamefont {Kenneth}\ \bibnamefont
  {Holmberg}}\ and\ \bibinfo {author} {\bibfnamefont {Ali}\ \bibnamefont
  {Erdemir}},\ }\bibfield  {title} {\enquote {\bibinfo {title} {Influence of
  tribology on global energy consumption, costs and emissions},}\ }\href
  {\doibase 10.1007/s40544-017-0183-5} {\bibfield  {journal} {\bibinfo
  {journal} {Friction}\ }\textbf {\bibinfo {volume} {5}},\ \bibinfo {pages}
  {263--284} (\bibinfo {year} {2017})}\BibitemShut {NoStop}%
\bibitem [{\citenamefont {Persson}(2000)}]{10.1007/978-3-662-04283-0}%
  \BibitemOpen
  \bibfield  {author} {\bibinfo {author} {\bibfnamefont {Bo~N.~J.}\
  \bibnamefont {Persson}},\ }\href {\doibase 10.1007/978-3-662-04283-0} {\emph
  {\bibinfo {title} {Sliding Friction}}}\ (\bibinfo  {publisher} {Springer
  Berlin Heidelberg},\ \bibinfo {year} {2000})\BibitemShut {NoStop}%
\bibitem [{\citenamefont {Persson}\ \emph {et~al.}(1999)\citenamefont
  {Persson}, \citenamefont {Tosatti}, \citenamefont {Fuhrmann}, \citenamefont
  {Witte},\ and\ \citenamefont {Wöll}}]{10.1103/physrevb.59.11777}%
  \BibitemOpen
  \bibfield  {author} {\bibinfo {author} {\bibfnamefont {B.~N.~J.}\
  \bibnamefont {Persson}}, \bibinfo {author} {\bibfnamefont {E.}~\bibnamefont
  {Tosatti}}, \bibinfo {author} {\bibfnamefont {D.}~\bibnamefont {Fuhrmann}},
  \bibinfo {author} {\bibfnamefont {G.}~\bibnamefont {Witte}}, \ and\ \bibinfo
  {author} {\bibfnamefont {Ch.}\ \bibnamefont {Wöll}},\ }\bibfield  {title}
  {\enquote {\bibinfo {title} {Low-frequency adsorbate vibrational relaxation
  and sliding friction},}\ }\href {\doibase 10.1103/physrevb.59.11777}
  {\bibfield  {journal} {\bibinfo  {journal} {Physical Review B}\ }\textbf
  {\bibinfo {volume} {59}},\ \bibinfo {pages} {11777--11791} (\bibinfo {year}
  {1999})}\BibitemShut {NoStop}%
\bibitem [{\citenamefont {Qi}\ \emph {et~al.}(2008)\citenamefont {Qi},
  \citenamefont {Park}, \citenamefont {Hendriksen}, \citenamefont {Ogletree},\
  and\ \citenamefont {Salmeron}}]{10.1103/physrevb.77.184105}%
  \BibitemOpen
  \bibfield  {author} {\bibinfo {author} {\bibfnamefont {Yabing}\ \bibnamefont
  {Qi}}, \bibinfo {author} {\bibfnamefont {J.~Y.}\ \bibnamefont {Park}},
  \bibinfo {author} {\bibfnamefont {B.~L.~M.}\ \bibnamefont {Hendriksen}},
  \bibinfo {author} {\bibfnamefont {D.~F.}\ \bibnamefont {Ogletree}}, \ and\
  \bibinfo {author} {\bibfnamefont {M.}~\bibnamefont {Salmeron}},\ }\bibfield
  {title} {\enquote {\bibinfo {title} {Electronic contribution to friction on
  {GaAs}: An atomic force microscope study},}\ }\href {\doibase
  10.1103/physrevb.77.184105} {\bibfield  {journal} {\bibinfo  {journal}
  {Physical Review B}\ }\textbf {\bibinfo {volume} {77}},\ \bibinfo {pages}
  {184105} (\bibinfo {year} {2008})}\BibitemShut {NoStop}%
\bibitem [{\citenamefont {Kadau}\ \emph {et~al.}(2008)\citenamefont {Kadau},
  \citenamefont {Hucht},\ and\ \citenamefont
  {Wolf}}]{10.1103/physrevlett.101.137205}%
  \BibitemOpen
  \bibfield  {author} {\bibinfo {author} {\bibfnamefont {Dirk}\ \bibnamefont
  {Kadau}}, \bibinfo {author} {\bibfnamefont {Alfred}\ \bibnamefont {Hucht}}, \
  and\ \bibinfo {author} {\bibfnamefont {Dietrich~E.}\ \bibnamefont {Wolf}},\
  }\bibfield  {title} {\enquote {\bibinfo {title} {Magnetic friction in ising
  spin systems},}\ }\href {\doibase 10.1103/physrevlett.101.137205} {\bibfield
  {journal} {\bibinfo  {journal} {Physical Review Letters}\ }\textbf {\bibinfo
  {volume} {101}},\ \bibinfo {pages} {137205} (\bibinfo {year}
  {2008})}\BibitemShut {NoStop}%
\bibitem [{\citenamefont {de~Wijn}\ \emph {et~al.}(2014)\citenamefont
  {de~Wijn}, \citenamefont {Fasolino}, \citenamefont {Filippov},\ and\
  \citenamefont {Urbakh}}]{10.1103/physrevlett.112.055502}%
  \BibitemOpen
  \bibfield  {author} {\bibinfo {author} {\bibfnamefont
  {A.{\hspace{0.167em}}S.}\ \bibnamefont {de~Wijn}}, \bibinfo {author}
  {\bibfnamefont {A.}~\bibnamefont {Fasolino}}, \bibinfo {author}
  {\bibfnamefont {A.{\hspace{0.167em}}E.}\ \bibnamefont {Filippov}}, \ and\
  \bibinfo {author} {\bibfnamefont {M.}~\bibnamefont {Urbakh}},\ }\bibfield
  {title} {\enquote {\bibinfo {title} {Nanoscopic friction under
  electrochemical control},}\ }\href {\doibase 10.1103/physrevlett.112.055502}
  {\bibfield  {journal} {\bibinfo  {journal} {Physical Review Letters}\
  }\textbf {\bibinfo {volume} {112}},\ \bibinfo {pages} {055502} (\bibinfo
  {year} {2014})}\BibitemShut {NoStop}%
\bibitem [{\citenamefont {Cieplak}\ \emph {et~al.}(1994)\citenamefont
  {Cieplak}, \citenamefont {Smith},\ and\ \citenamefont
  {Robbins}}]{10.1126/science.265.5176.1209}%
  \BibitemOpen
  \bibfield  {author} {\bibinfo {author} {\bibfnamefont {M.}~\bibnamefont
  {Cieplak}}, \bibinfo {author} {\bibfnamefont {E.~D.}\ \bibnamefont {Smith}},
  \ and\ \bibinfo {author} {\bibfnamefont {M.~O.}\ \bibnamefont {Robbins}},\
  }\bibfield  {title} {\enquote {\bibinfo {title} {Molecular origins of
  friction: The force on adsorbed layers},}\ }\href {\doibase
  10.1126/science.265.5176.1209} {\bibfield  {journal} {\bibinfo  {journal}
  {Science}\ }\textbf {\bibinfo {volume} {265}},\ \bibinfo {pages} {1209--1212}
  (\bibinfo {year} {1994})}\BibitemShut {NoStop}%
\bibitem [{\citenamefont {Hu}\ \emph {et~al.}(2019)\citenamefont {Hu},
  \citenamefont {Krylov},\ and\ \citenamefont
  {Frenken}}]{10.1007/s11249-019-1247-7}%
  \BibitemOpen
  \bibfield  {author} {\bibinfo {author} {\bibfnamefont {Renfeng}\ \bibnamefont
  {Hu}}, \bibinfo {author} {\bibfnamefont {Sergey~Yu.}\ \bibnamefont {Krylov}},
  \ and\ \bibinfo {author} {\bibfnamefont {Joost W.~M.}\ \bibnamefont
  {Frenken}},\ }\bibfield  {title} {\enquote {\bibinfo {title} {On the origin
  of frictional energy dissipation},}\ }\href {\doibase
  10.1007/s11249-019-1247-7} {\bibfield  {journal} {\bibinfo  {journal}
  {Tribology Letters}\ }\textbf {\bibinfo {volume} {68}} (\bibinfo {year}
  {2019}),\ 10.1007/s11249-019-1247-7}\BibitemShut {NoStop}%
\bibitem [{\citenamefont {Smith}\ \emph {et~al.}(1996)\citenamefont {Smith},
  \citenamefont {Robbins},\ and\ \citenamefont
  {Cieplak}}]{10.1103/physrevb.54.8252}%
  \BibitemOpen
  \bibfield  {author} {\bibinfo {author} {\bibfnamefont {Elizabeth~D.}\
  \bibnamefont {Smith}}, \bibinfo {author} {\bibfnamefont {Mark~O.}\
  \bibnamefont {Robbins}}, \ and\ \bibinfo {author} {\bibfnamefont {Marek}\
  \bibnamefont {Cieplak}},\ }\bibfield  {title} {\enquote {\bibinfo {title}
  {Friction on adsorbed monolayers},}\ }\href {\doibase
  10.1103/physrevb.54.8252} {\bibfield  {journal} {\bibinfo  {journal}
  {Physical Review B}\ }\textbf {\bibinfo {volume} {54}},\ \bibinfo {pages}
  {8252--8260} (\bibinfo {year} {1996})}\BibitemShut {NoStop}%
\bibitem [{\citenamefont {Persson}(2020)}]{10.1007/s11249-020-1268-2}%
  \BibitemOpen
  \bibfield  {author} {\bibinfo {author} {\bibfnamefont {B.~N.~J.}\
  \bibnamefont {Persson}},\ }\bibfield  {title} {\enquote {\bibinfo {title}
  {Comment on {\textquotedblleft}on the origin of frictional energy
  dissipation{\textquotedblright}},}\ }\href {\doibase
  10.1007/s11249-020-1268-2} {\bibfield  {journal} {\bibinfo  {journal}
  {Tribology Letters}\ }\textbf {\bibinfo {volume} {68}} (\bibinfo {year}
  {2020}),\ 10.1007/s11249-020-1268-2}\BibitemShut {NoStop}%
\bibitem [{\citenamefont {Hu}\ \emph {et~al.}(2020)\citenamefont {Hu},
  \citenamefont {Krylov},\ and\ \citenamefont
  {Frenken}}]{10.1007/s11249-020-1280-6}%
  \BibitemOpen
  \bibfield  {author} {\bibinfo {author} {\bibfnamefont {Renfeng}\ \bibnamefont
  {Hu}}, \bibinfo {author} {\bibfnamefont {Sergey~Yu.}\ \bibnamefont {Krylov}},
  \ and\ \bibinfo {author} {\bibfnamefont {Joost W.~M.}\ \bibnamefont
  {Frenken}},\ }\bibfield  {title} {\enquote {\bibinfo {title} {Response to
  comment on {\textquotedblleft}on the origin of frictional energy
  dissipation{\textquotedblright}, by b.n.j. persson},}\ }\href {\doibase
  10.1007/s11249-020-1280-6} {\bibfield  {journal} {\bibinfo  {journal}
  {Tribology Letters}\ }\textbf {\bibinfo {volume} {68}} (\bibinfo {year}
  {2020}),\ 10.1007/s11249-020-1280-6}\BibitemShut {NoStop}%
\bibitem [{\citenamefont {Wei}\ \emph {et~al.}(2020)\citenamefont {Wei},
  \citenamefont {Duan}, \citenamefont {Kan}, \citenamefont {Zhang},\ and\
  \citenamefont {Chen}}]{10.1063/1.5130705}%
  \BibitemOpen
  \bibfield  {author} {\bibinfo {author} {\bibfnamefont {Zhiyong}\ \bibnamefont
  {Wei}}, \bibinfo {author} {\bibfnamefont {Zaoqi}\ \bibnamefont {Duan}},
  \bibinfo {author} {\bibfnamefont {Yajing}\ \bibnamefont {Kan}}, \bibinfo
  {author} {\bibfnamefont {Yan}\ \bibnamefont {Zhang}}, \ and\ \bibinfo
  {author} {\bibfnamefont {Yunfei}\ \bibnamefont {Chen}},\ }\bibfield  {title}
  {\enquote {\bibinfo {title} {Phonon energy dissipation in friction between
  graphene/graphene interface},}\ }\href {\doibase 10.1063/1.5130705}
  {\bibfield  {journal} {\bibinfo  {journal} {Journal of Applied Physics}\
  }\textbf {\bibinfo {volume} {127}},\ \bibinfo {pages} {015105} (\bibinfo
  {year} {2020})}\BibitemShut {NoStop}%
\bibitem [{\citenamefont {Braun}\ and\ \citenamefont
  {Kivshar}(2004)}]{10.1007/978-3-662-10331-9}%
  \BibitemOpen
  \bibfield  {author} {\bibinfo {author} {\bibfnamefont {Oleg~M.}\ \bibnamefont
  {Braun}}\ and\ \bibinfo {author} {\bibfnamefont {Yuri~S.}\ \bibnamefont
  {Kivshar}},\ }\href {\doibase 10.1007/978-3-662-10331-9} {\emph {\bibinfo
  {title} {The Frenkel-Kontorova Model}}}\ (\bibinfo  {publisher} {Springer
  Berlin Heidelberg},\ \bibinfo {year} {2004})\BibitemShut {NoStop}%
\bibitem [{\citenamefont {Norell}\ \emph {et~al.}(2016)\citenamefont {Norell},
  \citenamefont {Fasolino},\ and\ \citenamefont
  {de~Wijn}}]{10.1103/physreve.94.023001}%
  \BibitemOpen
  \bibfield  {author} {\bibinfo {author} {\bibfnamefont {Jesper}\ \bibnamefont
  {Norell}}, \bibinfo {author} {\bibfnamefont {Annalisa}\ \bibnamefont
  {Fasolino}}, \ and\ \bibinfo {author} {\bibfnamefont {Astrid~S.}\
  \bibnamefont {de~Wijn}},\ }\bibfield  {title} {\enquote {\bibinfo {title}
  {Emergent friction in two-dimensional frenkel-kontorova models},}\ }\href
  {\doibase 10.1103/physreve.94.023001} {\bibfield  {journal} {\bibinfo
  {journal} {Physical Review E}\ }\textbf {\bibinfo {volume} {94}},\ \bibinfo
  {pages} {023001} (\bibinfo {year} {2016})}\BibitemShut {NoStop}%
\bibitem [{Note1()}]{Note1}%
  \BibitemOpen
  \bibinfo {note} {See Fig.~8 of Ref.~\cite {10.1103/physreve.94.023001}. The
  ``vector'' model variant, in which all higher-order terms in the particle
  interactions are retained, thus making it anharmonic, displays significantly
  higher friction.}\BibitemShut {Stop}%
\bibitem [{\citenamefont {Vink}(2019)}]{10.1103/physrevb.100.094305}%
  \BibitemOpen
  \bibfield  {author} {\bibinfo {author} {\bibfnamefont {Richard L.~C.}\
  \bibnamefont {Vink}},\ }\bibfield  {title} {\enquote {\bibinfo {title}
  {Connection between sliding friction and phonon lifetimes: Thermostat-induced
  thermolubricity effects in molecular dynamics simulations},}\ }\href
  {\doibase 10.1103/physrevb.100.094305} {\bibfield  {journal} {\bibinfo
  {journal} {Physical Review B}\ }\textbf {\bibinfo {volume} {100}},\ \bibinfo
  {pages} {094305} (\bibinfo {year} {2019})}\BibitemShut {NoStop}%
\bibitem [{\citenamefont {Schmidt}\ \emph {et~al.}(2020)\citenamefont
  {Schmidt}, \citenamefont {Krisponeit}, \citenamefont {Weber}, \citenamefont
  {Samwer},\ and\ \citenamefont {Volkert}}]{schmidt2020switching}%
  \BibitemOpen
  \bibfield  {author} {\bibinfo {author} {\bibfnamefont {H.}~\bibnamefont
  {Schmidt}}, \bibinfo {author} {\bibfnamefont {J.~O.}\ \bibnamefont
  {Krisponeit}}, \bibinfo {author} {\bibfnamefont {N.}~\bibnamefont {Weber}},
  \bibinfo {author} {\bibfnamefont {K.}~\bibnamefont {Samwer}}, \ and\ \bibinfo
  {author} {\bibfnamefont {C.~A.}\ \bibnamefont {Volkert}},\ }\href@noop {}
  {\enquote {\bibinfo {title} {Switching friction at a manganite surface using
  electric fields},}\ } (\bibinfo {year} {2020}),\ \Eprint
  {http://arxiv.org/abs/2005.08949} {arXiv:2005.08949 [cond-mat.mtrl-sci]}
  \BibitemShut {NoStop}%
\bibitem [{\citenamefont {Weber}\ \emph {et~al.}(2020)\citenamefont {Weber},
  \citenamefont {Schmidt}, \citenamefont {Sievert}, \citenamefont {Jooss},
  \citenamefont {Güthoff}, \citenamefont {Moshneaga}, \citenamefont {Samwer},
  \citenamefont {Krüger},\ and\ \citenamefont {Volkert}}]{weber2020polaronic}%
  \BibitemOpen
  \bibfield  {author} {\bibinfo {author} {\bibfnamefont {Niklas~A.}\
  \bibnamefont {Weber}}, \bibinfo {author} {\bibfnamefont {Dr.~Hendrik}\
  \bibnamefont {Schmidt}}, \bibinfo {author} {\bibfnamefont {Tim}\ \bibnamefont
  {Sievert}}, \bibinfo {author} {\bibfnamefont {Prof.~Christian}\ \bibnamefont
  {Jooss}}, \bibinfo {author} {\bibfnamefont {Dr.~Friedrich}\ \bibnamefont
  {Güthoff}}, \bibinfo {author} {\bibfnamefont {Prof.~Vasily}\ \bibnamefont
  {Moshneaga}}, \bibinfo {author} {\bibfnamefont {Prof.~Konrad}\ \bibnamefont
  {Samwer}}, \bibinfo {author} {\bibfnamefont {Prof.~Matthias}\ \bibnamefont
  {Krüger}}, \ and\ \bibinfo {author} {\bibfnamefont {Prof. Cynthia~A.}\
  \bibnamefont {Volkert}},\ }\href@noop {} {\enquote {\bibinfo {title}
  {Polaronic contributions to friction in a manganite thin film},}\ } (\bibinfo
  {year} {2020}),\ \Eprint {http://arxiv.org/abs/2009.12137} {arXiv:2009.12137
  [cond-mat.mtrl-sci]} \BibitemShut {NoStop}%
\bibitem [{\citenamefont {Consoli}\ \emph {et~al.}(2000)\citenamefont
  {Consoli}, \citenamefont {Knops},\ and\ \citenamefont
  {Fasolino}}]{10.1103/physrevlett.85.302}%
  \BibitemOpen
  \bibfield  {author} {\bibinfo {author} {\bibfnamefont {L.}~\bibnamefont
  {Consoli}}, \bibinfo {author} {\bibfnamefont {H.~J.~F.}\ \bibnamefont
  {Knops}}, \ and\ \bibinfo {author} {\bibfnamefont {A.}~\bibnamefont
  {Fasolino}},\ }\bibfield  {title} {\enquote {\bibinfo {title} {Onset of
  sliding friction in incommensurate systems},}\ }\href {\doibase
  10.1103/physrevlett.85.302} {\bibfield  {journal} {\bibinfo  {journal}
  {Physical Review Letters}\ }\textbf {\bibinfo {volume} {85}},\ \bibinfo
  {pages} {302--305} (\bibinfo {year} {2000})}\BibitemShut {NoStop}%
\bibitem [{\citenamefont {Consoli}\ \emph {et~al.}(2001)\citenamefont
  {Consoli}, \citenamefont {Knops},\ and\ \citenamefont
  {Fasolino}}]{10.1103/physreve.64.016601}%
  \BibitemOpen
  \bibfield  {author} {\bibinfo {author} {\bibfnamefont {L.}~\bibnamefont
  {Consoli}}, \bibinfo {author} {\bibfnamefont {H.~J.~F.}\ \bibnamefont
  {Knops}}, \ and\ \bibinfo {author} {\bibfnamefont {A.}~\bibnamefont
  {Fasolino}},\ }\bibfield  {title} {\enquote {\bibinfo {title} {Breakdown of a
  conservation law in incommensurate systems},}\ }\href {\doibase
  10.1103/physreve.64.016601} {\bibfield  {journal} {\bibinfo  {journal}
  {Physical Review E}\ }\textbf {\bibinfo {volume} {64}},\ \bibinfo {pages}
  {016601} (\bibinfo {year} {2001})}\BibitemShut {NoStop}%
\bibitem [{\citenamefont {Peyrard}\ and\ \citenamefont
  {Aubry}(1983)}]{10.1088/0022-3719/16/9/005}%
  \BibitemOpen
  \bibfield  {author} {\bibinfo {author} {\bibfnamefont {M}~\bibnamefont
  {Peyrard}}\ and\ \bibinfo {author} {\bibfnamefont {S}~\bibnamefont {Aubry}},\
  }\bibfield  {title} {\enquote {\bibinfo {title} {Critical behaviour at the
  transition by breaking of analyticity in the discrete frenkel-kontorova
  model},}\ }\href {\doibase 10.1088/0022-3719/16/9/005} {\bibfield  {journal}
  {\bibinfo  {journal} {Journal of Physics C: Solid State Physics}\ }\textbf
  {\bibinfo {volume} {16}},\ \bibinfo {pages} {1593--1608} (\bibinfo {year}
  {1983})}\BibitemShut {NoStop}%
\bibitem [{\citenamefont {Shinjo}\ and\ \citenamefont
  {Hirano}(1993)}]{10.1016/0039-6028(93)91022-h}%
  \BibitemOpen
  \bibfield  {author} {\bibinfo {author} {\bibfnamefont {Kazumasa}\
  \bibnamefont {Shinjo}}\ and\ \bibinfo {author} {\bibfnamefont {Motohisa}\
  \bibnamefont {Hirano}},\ }\bibfield  {title} {\enquote {\bibinfo {title}
  {Dynamics of friction: superlubric state},}\ }\href {\doibase
  10.1016/0039-6028(93)91022-h} {\bibfield  {journal} {\bibinfo  {journal}
  {Surface Science}\ }\textbf {\bibinfo {volume} {283}},\ \bibinfo {pages}
  {473--478} (\bibinfo {year} {1993})}\BibitemShut {NoStop}%
\bibitem [{\citenamefont {Bylinskii}\ \emph {et~al.}(2016)\citenamefont
  {Bylinskii}, \citenamefont {Gangloff}, \citenamefont {Counts},\ and\
  \citenamefont {Vuleti{\'{c}}}}]{10.1038/nmat4601}%
  \BibitemOpen
  \bibfield  {author} {\bibinfo {author} {\bibfnamefont {Alexei}\ \bibnamefont
  {Bylinskii}}, \bibinfo {author} {\bibfnamefont {Dorian}\ \bibnamefont
  {Gangloff}}, \bibinfo {author} {\bibfnamefont {Ian}\ \bibnamefont {Counts}},
  \ and\ \bibinfo {author} {\bibfnamefont {Vladan}\ \bibnamefont
  {Vuleti{\'{c}}}},\ }\bibfield  {title} {\enquote {\bibinfo {title}
  {Observation of aubry-type transition in finite atom chains via friction},}\
  }\href {\doibase 10.1038/nmat4601} {\bibfield  {journal} {\bibinfo  {journal}
  {Nature Materials}\ }\textbf {\bibinfo {volume} {15}},\ \bibinfo {pages}
  {717--721} (\bibinfo {year} {2016})}\BibitemShut {NoStop}%
\bibitem [{\citenamefont {van~den Ende}\ \emph {et~al.}(2012)\citenamefont
  {van~den Ende}, \citenamefont {de~Wijn},\ and\ \citenamefont
  {Fasolino}}]{10.1088/0953-8984/24/44/445009}%
  \BibitemOpen
  \bibfield  {author} {\bibinfo {author} {\bibfnamefont {Joost~A}\ \bibnamefont
  {van~den Ende}}, \bibinfo {author} {\bibfnamefont {Astrid~S}\ \bibnamefont
  {de~Wijn}}, \ and\ \bibinfo {author} {\bibfnamefont {Annalisa}\ \bibnamefont
  {Fasolino}},\ }\bibfield  {title} {\enquote {\bibinfo {title} {The effect of
  temperature and velocity on superlubricity},}\ }\href {\doibase
  10.1088/0953-8984/24/44/445009} {\bibfield  {journal} {\bibinfo  {journal}
  {Journal of Physics: Condensed Matter}\ }\textbf {\bibinfo {volume} {24}},\
  \bibinfo {pages} {445009} (\bibinfo {year} {2012})}\BibitemShut {NoStop}%
\bibitem [{\citenamefont {Dienwiebel}\ \emph {et~al.}(2004)\citenamefont
  {Dienwiebel}, \citenamefont {Verhoeven}, \citenamefont {Pradeep},
  \citenamefont {Frenken}, \citenamefont {Heimberg},\ and\ \citenamefont
  {Zandbergen}}]{10.1103/physrevlett.92.126101}%
  \BibitemOpen
  \bibfield  {author} {\bibinfo {author} {\bibfnamefont {Martin}\ \bibnamefont
  {Dienwiebel}}, \bibinfo {author} {\bibfnamefont {Gertjan~S.}\ \bibnamefont
  {Verhoeven}}, \bibinfo {author} {\bibfnamefont {Namboodiri}\ \bibnamefont
  {Pradeep}}, \bibinfo {author} {\bibfnamefont {Joost W.~M.}\ \bibnamefont
  {Frenken}}, \bibinfo {author} {\bibfnamefont {Jennifer~A.}\ \bibnamefont
  {Heimberg}}, \ and\ \bibinfo {author} {\bibfnamefont {Henny~W.}\ \bibnamefont
  {Zandbergen}},\ }\bibfield  {title} {\enquote {\bibinfo {title}
  {Superlubricity of graphite},}\ }\href {\doibase
  10.1103/physrevlett.92.126101} {\bibfield  {journal} {\bibinfo  {journal}
  {Physical Review Letters}\ }\textbf {\bibinfo {volume} {92}},\ \bibinfo
  {pages} {126101} (\bibinfo {year} {2004})}\BibitemShut {NoStop}%
\bibitem [{\citenamefont {de~Wijn}\ \emph {et~al.}(2011)\citenamefont
  {de~Wijn}, \citenamefont {Fasolino}, \citenamefont {Filippov},\ and\
  \citenamefont {Urbakh}}]{10.1209/0295-5075/95/66002}%
  \BibitemOpen
  \bibfield  {author} {\bibinfo {author} {\bibfnamefont {Astrid~S.}\
  \bibnamefont {de~Wijn}}, \bibinfo {author} {\bibfnamefont {Annalisa}\
  \bibnamefont {Fasolino}}, \bibinfo {author} {\bibfnamefont {A.~E.}\
  \bibnamefont {Filippov}}, \ and\ \bibinfo {author} {\bibfnamefont
  {M.}~\bibnamefont {Urbakh}},\ }\bibfield  {title} {\enquote {\bibinfo {title}
  {Low friction and rotational dynamics of crystalline flakes in solid
  lubrication},}\ }\href {\doibase 10.1209/0295-5075/95/66002} {\bibfield
  {journal} {\bibinfo  {journal} {{EPL} (Europhysics Letters)}\ }\textbf
  {\bibinfo {volume} {95}},\ \bibinfo {pages} {66002} (\bibinfo {year}
  {2011})}\BibitemShut {NoStop}%
\bibitem [{\citenamefont {Tomassone}\ \emph {et~al.}(1997)\citenamefont
  {Tomassone}, \citenamefont {Sokoloff}, \citenamefont {Widom},\ and\
  \citenamefont {Krim}}]{10.1103/physrevlett.79.4798}%
  \BibitemOpen
  \bibfield  {author} {\bibinfo {author} {\bibfnamefont {M.~S.}\ \bibnamefont
  {Tomassone}}, \bibinfo {author} {\bibfnamefont {J.~B.}\ \bibnamefont
  {Sokoloff}}, \bibinfo {author} {\bibfnamefont {A.}~\bibnamefont {Widom}}, \
  and\ \bibinfo {author} {\bibfnamefont {J.}~\bibnamefont {Krim}},\ }\bibfield
  {title} {\enquote {\bibinfo {title} {Dominance of phonon friction for a xenon
  film on a silver (111) surface},}\ }\href {\doibase
  10.1103/physrevlett.79.4798} {\bibfield  {journal} {\bibinfo  {journal}
  {Physical Review Letters}\ }\textbf {\bibinfo {volume} {79}},\ \bibinfo
  {pages} {4798--4801} (\bibinfo {year} {1997})}\BibitemShut {NoStop}%
\bibitem [{\citenamefont {Benassi}\ \emph {et~al.}(2010)\citenamefont
  {Benassi}, \citenamefont {Vanossi}, \citenamefont {Santoro},\ and\
  \citenamefont {Tosatti}}]{10.1103/physrevb.82.081401}%
  \BibitemOpen
  \bibfield  {author} {\bibinfo {author} {\bibfnamefont {A.}~\bibnamefont
  {Benassi}}, \bibinfo {author} {\bibfnamefont {A.}~\bibnamefont {Vanossi}},
  \bibinfo {author} {\bibfnamefont {G.~E.}\ \bibnamefont {Santoro}}, \ and\
  \bibinfo {author} {\bibfnamefont {E.}~\bibnamefont {Tosatti}},\ }\bibfield
  {title} {\enquote {\bibinfo {title} {Parameter-free dissipation in simulated
  sliding friction},}\ }\href {\doibase 10.1103/physrevb.82.081401} {\bibfield
  {journal} {\bibinfo  {journal} {Physical Review B}\ }\textbf {\bibinfo
  {volume} {82}},\ \bibinfo {pages} {081401} (\bibinfo {year}
  {2010})}\BibitemShut {NoStop}%
\bibitem [{\citenamefont {Benassi}\ \emph {et~al.}(2012)\citenamefont
  {Benassi}, \citenamefont {Vanossi}, \citenamefont {Santoro},\ and\
  \citenamefont {Tosatti}}]{10.1007/s11249-012-9936-5}%
  \BibitemOpen
  \bibfield  {author} {\bibinfo {author} {\bibfnamefont {A.}~\bibnamefont
  {Benassi}}, \bibinfo {author} {\bibfnamefont {A.}~\bibnamefont {Vanossi}},
  \bibinfo {author} {\bibfnamefont {G.~E.}\ \bibnamefont {Santoro}}, \ and\
  \bibinfo {author} {\bibfnamefont {E.}~\bibnamefont {Tosatti}},\ }\bibfield
  {title} {\enquote {\bibinfo {title} {Optimal energy dissipation in sliding
  friction simulations},}\ }\href {\doibase 10.1007/s11249-012-9936-5}
  {\bibfield  {journal} {\bibinfo  {journal} {Tribology Letters}\ }\textbf
  {\bibinfo {volume} {48}},\ \bibinfo {pages} {41--49} (\bibinfo {year}
  {2012})}\BibitemShut {NoStop}%
\bibitem [{\citenamefont {Lebedeva}\ \emph {et~al.}(2011)\citenamefont
  {Lebedeva}, \citenamefont {Knizhnik}, \citenamefont {Popov}, \citenamefont
  {Lozovik},\ and\ \citenamefont {Potapkin}}]{10.1039/c0cp02614j}%
  \BibitemOpen
  \bibfield  {author} {\bibinfo {author} {\bibfnamefont {Irina~V.}\
  \bibnamefont {Lebedeva}}, \bibinfo {author} {\bibfnamefont {Andrey~A.}\
  \bibnamefont {Knizhnik}}, \bibinfo {author} {\bibfnamefont {Andrey~M.}\
  \bibnamefont {Popov}}, \bibinfo {author} {\bibfnamefont {Yurii~E.}\
  \bibnamefont {Lozovik}}, \ and\ \bibinfo {author} {\bibfnamefont {Boris~V.}\
  \bibnamefont {Potapkin}},\ }\bibfield  {title} {\enquote {\bibinfo {title}
  {Interlayer interaction and relative vibrations of bilayer graphene},}\
  }\href {\doibase 10.1039/c0cp02614j} {\bibfield  {journal} {\bibinfo
  {journal} {Physical Chemistry Chemical Physics}\ }\textbf {\bibinfo {volume}
  {13}},\ \bibinfo {pages} {5687} (\bibinfo {year} {2011})}\BibitemShut
  {NoStop}%
\bibitem [{\citenamefont {Apostoli}\ \emph {et~al.}(2017)\citenamefont
  {Apostoli}, \citenamefont {Giusti}, \citenamefont {Ciccoianni}, \citenamefont
  {Riva}, \citenamefont {Capozza}, \citenamefont {Woulach{\'{e}}},
  \citenamefont {Vanossi}, \citenamefont {Panizon},\ and\ \citenamefont
  {Manini}}]{10.3762/bjnano.8.218}%
  \BibitemOpen
  \bibfield  {author} {\bibinfo {author} {\bibfnamefont {Christian}\
  \bibnamefont {Apostoli}}, \bibinfo {author} {\bibfnamefont {Giovanni}\
  \bibnamefont {Giusti}}, \bibinfo {author} {\bibfnamefont {Jacopo}\
  \bibnamefont {Ciccoianni}}, \bibinfo {author} {\bibfnamefont {Gabriele}\
  \bibnamefont {Riva}}, \bibinfo {author} {\bibfnamefont {Rosario}\
  \bibnamefont {Capozza}}, \bibinfo {author} {\bibfnamefont {Rosalie~Laure}\
  \bibnamefont {Woulach{\'{e}}}}, \bibinfo {author} {\bibfnamefont {Andrea}\
  \bibnamefont {Vanossi}}, \bibinfo {author} {\bibfnamefont {Emanuele}\
  \bibnamefont {Panizon}}, \ and\ \bibinfo {author} {\bibfnamefont {Nicola}\
  \bibnamefont {Manini}},\ }\bibfield  {title} {\enquote {\bibinfo {title}
  {Velocity dependence of sliding friction on a crystalline surface},}\ }\href
  {\doibase 10.3762/bjnano.8.218} {\bibfield  {journal} {\bibinfo  {journal}
  {Beilstein Journal of Nanotechnology}\ }\textbf {\bibinfo {volume} {8}},\
  \bibinfo {pages} {2186--2199} (\bibinfo {year} {2017})}\BibitemShut {NoStop}%
\bibitem [{\citenamefont {Plimpton}(1995)}]{10.1006/jcph.1995.1039}%
  \BibitemOpen
  \bibfield  {author} {\bibinfo {author} {\bibfnamefont {Steve}\ \bibnamefont
  {Plimpton}},\ }\bibfield  {title} {\enquote {\bibinfo {title} {Fast parallel
  algorithms for short-range molecular dynamics},}\ }\href {\doibase
  10.1006/jcph.1995.1039} {\bibfield  {journal} {\bibinfo  {journal} {Journal
  of Computational Physics}\ }\textbf {\bibinfo {volume} {117}},\ \bibinfo
  {pages} {1--19} (\bibinfo {year} {1995})}\BibitemShut {NoStop}%
\bibitem [{\citenamefont {van Erp}\ \emph {et~al.}(1999)\citenamefont {van
  Erp}, \citenamefont {Fasolino}, \citenamefont {Radulescu},\ and\
  \citenamefont {Janssen}}]{10.1103/physrevb.60.6522}%
  \BibitemOpen
  \bibfield  {author} {\bibinfo {author} {\bibfnamefont {T.~S.}\ \bibnamefont
  {van Erp}}, \bibinfo {author} {\bibfnamefont {A.}~\bibnamefont {Fasolino}},
  \bibinfo {author} {\bibfnamefont {O.}~\bibnamefont {Radulescu}}, \ and\
  \bibinfo {author} {\bibfnamefont {T.}~\bibnamefont {Janssen}},\ }\bibfield
  {title} {\enquote {\bibinfo {title} {Pinning and phonon localization in
  frenkel-kontorova models on quasiperiodic substrates},}\ }\href {\doibase
  10.1103/physrevb.60.6522} {\bibfield  {journal} {\bibinfo  {journal}
  {Physical Review B}\ }\textbf {\bibinfo {volume} {60}},\ \bibinfo {pages}
  {6522--6528} (\bibinfo {year} {1999})}\BibitemShut {NoStop}%
\bibitem [{\citenamefont {Varini}\ \emph {et~al.}(2015)\citenamefont {Varini},
  \citenamefont {Vanossi}, \citenamefont {Guerra}, \citenamefont {Mandelli},
  \citenamefont {Capozza},\ and\ \citenamefont {Tosatti}}]{10.1039/c4nr06521b}%
  \BibitemOpen
  \bibfield  {author} {\bibinfo {author} {\bibfnamefont {Nicola}\ \bibnamefont
  {Varini}}, \bibinfo {author} {\bibfnamefont {Andrea}\ \bibnamefont
  {Vanossi}}, \bibinfo {author} {\bibfnamefont {Roberto}\ \bibnamefont
  {Guerra}}, \bibinfo {author} {\bibfnamefont {Davide}\ \bibnamefont
  {Mandelli}}, \bibinfo {author} {\bibfnamefont {Rosario}\ \bibnamefont
  {Capozza}}, \ and\ \bibinfo {author} {\bibfnamefont {Erio}\ \bibnamefont
  {Tosatti}},\ }\bibfield  {title} {\enquote {\bibinfo {title} {Static friction
  scaling of physisorbed islands: the key is in the edge},}\ }\href {\doibase
  10.1039/c4nr06521b} {\bibfield  {journal} {\bibinfo  {journal} {Nanoscale}\
  }\textbf {\bibinfo {volume} {7}},\ \bibinfo {pages} {2093--2101} (\bibinfo
  {year} {2015})}\BibitemShut {NoStop}%
\bibitem [{\citenamefont {Cserti}\ and\ \citenamefont
  {Tichy}(2004)}]{10.1088/0143-0807/25/6/004}%
  \BibitemOpen
  \bibfield  {author} {\bibinfo {author} {\bibfnamefont {J{\'{o}}zsef}\
  \bibnamefont {Cserti}}\ and\ \bibinfo {author} {\bibfnamefont {G{\'{e}}za}\
  \bibnamefont {Tichy}},\ }\bibfield  {title} {\enquote {\bibinfo {title} {A
  simple model for the vibrational modes in honeycomb lattices},}\ }\href
  {\doibase 10.1088/0143-0807/25/6/004} {\bibfield  {journal} {\bibinfo
  {journal} {European Journal of Physics}\ }\textbf {\bibinfo {volume} {25}},\
  \bibinfo {pages} {723--736} (\bibinfo {year} {2004})}\BibitemShut {NoStop}%
\end{thebibliography}%

\appendix

\section{MD implementation details}

Both the 1D and 2D model can be implemented in LAMMPS~\cite{10.1006/jcph.1995.1039} using mainly standard features (the exception is the anharmonic bond potential of \eq{eq:spr}, for which we modified an existing bond style). All data were obtained in the microcanonical ensemble (\texttt{fix~nve}, timestep~0.001). For the rigid flake, \texttt{fix~rigid/nve} was used to implement the rigid edge. The 
essential simulation output is the trajectory, i.e.~particle positions and 
velocities as a function of time, from which all quantities of interest can be 
computed. For the computation of the hessian and eigenmodes, a dedicated C-code 
was used, based on LAPACK. This code was also used to process the MD trajectory, 
in order to obtain the mode kinetic energy and amplitude.

\section{Coherent distribution}

In the strict absence of phonon scattering, the quantity $\vec{v}_i (t) \cdot 
\vec{\xi}_{k,i}$ in \eq{eq:kin} is a periodic function at the eigenfrequency 
$\omega_k$, implying for the kinetic energy $K_k (t) = K_{0,k} \cos^2 (\omega_k 
t + \phi_k)$, with amplitude $K_{0,k}$, and phase $\phi_k$. Converting the time 
series signal $K_k(t)$ into a histogram, the inverse height of the bin 
corresponding to the energy value $K_k$ will be given by
\begin{equation}
 \frac{1}{H(K_k)} \propto \left| \frac{dK_k}{dt} \right|
 \propto \sqrt{K_k(K_{0,k}-K_k)} \approx c \sqrt{K_k},
\end{equation}
with $c$ a constant, and where the approximation refers to the limit of small 
$K_k$, which \eq{eq:pk} uses. If one does not make this approximation, then the 
histogram $H(K_k)$ will actually reveal two peaks, at $K_k=0,K_{0,k}$. For the 
1D chain {\it without} the external field, which then is a true harmonic system 
where phonon scattering is strictly absent, this is indeed what one observes. 
However, in the presence of the external field (induced by the static 
obstacles), we never observed the second high-energy peak, since this peak is 
then exponentially suppressed by the Maxwell-Boltzmann factor.

\section{Dispersion relation hexagonal lattice}

\begin{figure}[t]
\begin{center}
\includegraphics[width=\columnwidth]{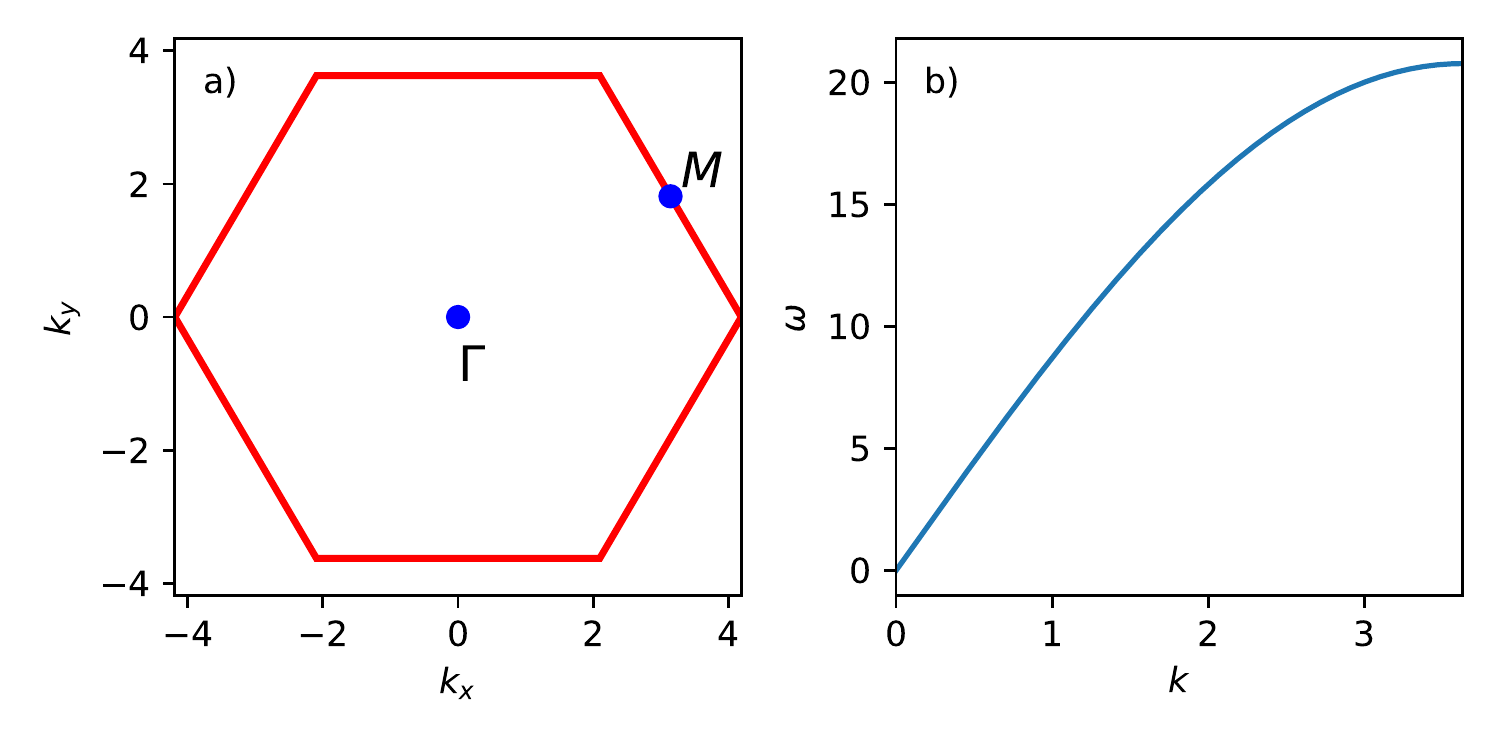}
\caption{\label{fig9} a) FBZ of the hexagonal lattice. b) Longitudinal dispersion along the line $\Gamma M$.}
\end{center}
\end{figure}

With the hexagonal sliding lattice oriented in the $(xy)$-coordinate system as shown in \fig{fig:2d_sketch}(a), the first Brillouin zone (FBZ) is a hexagon oriented as shown in \fig{fig9}(a), where $\Gamma M = 2\pi / (\sqrt{3} \, a)$ indicates the $+30^{\rm o}$ propagation direction. \fig{fig9}(b) shows the longitudinal dispersion along $\Gamma M$, with $\omega$ expressed in the units of our model. The dispersion relation was computed numerically using equations provided in \olcite{10.1088/0143-0807/25/6/004}. For values of $k$ outside the interval $\Gamma M$, one uses the {\it periodic even extension} of the dispersion relation to obtain the frequency.

\end{document}